\documentclass[a4paper,12pt,final]{article}
\usepackage{color}
\usepackage{bbm}
\usepackage{amssymb,amsmath,amssymb,amsthm}
\usepackage{pdfpages}
\usepackage{pgfplots}
\usepackage{tikz}
\usetikzlibrary{matrix,arrows,snakes}
\usepackage{array,multirow}
\usepackage{hyperref} 
\usepackage[notcite,notref]{showkeys}
\usepackage{bm}
\usepackage{todonotes}
\usepackage{graphicx}

\usepackage[T1]{fontenc}
\usepackage[utf8]{inputenc}
\usepackage{authblk}

\usepackage{epstopdf}
\usepackage{amsfonts,mathrsfs,array,pifont}

\newtheorem{proposition}{Proposition}

\renewcommand\marginpar[1]{}

\def\be{\begin{equation}}
\def\ee{\end{equation}}
\def\ba{\begin{array}}
\def\ea{\end{array}}

\newcommand\comment[1]{}

\definecolor{red}{rgb}{0, 0, 0}
\definecolor{blue}{rgb}{0, 0, 0}
\definecolor{blue2}{rgb}{0.0, 0.0, 0.0}

\title{Steady accretion of an elastic body on a hard spherical surface and the notion of a four-dimensional reference space}

\author[a]{Giuseppe Tomassetti}
\author[b]{Tal Cohen}
\author[c]{Rohan Abeyaratne}

\affil[a]{DICII Department, University of Rome Tor Vergata, Italy. Email: \texttt{tomassetti@ing.uniroma2.it}}

\affil[b]{John A. Paulson School of Engineering and Applied Sciences, Harvard University, Cambridge, MA, USA. Email: \texttt{talcohen@fas.harvard.edu}}

\affil[c]{Department of Mechanical Engineering, Massachusetts Institute of Technology, Cambridge, MA, USA.  Email: \texttt{rohan@mit.edu}}

\begin{document}

\maketitle

\begin{abstract}

Taking the cue from experiments on actin growth on spherical beads, we formulate and solve a model problem describing the accretion of an incompressible elastic solid on a rigid sphere due to attachment of diffusing free particles. One of the peculiar characteristics of  this problem is that accretion takes place on the interior surface that separates the body from its support rather than on its exterior surface, and hence is responsible for stress accumulation. Simultaneously, ablation takes place at the outer surface where material is removed from the body. As the body grows, mechanical effects associated with the build-up of stress and strain energy slow down accretion and promote ablation. Eventually, the system reaches a point where internal accretion is balanced by external ablation. The present study is concerned with this stationary regime called ``treadmilling''.

The principal ingredients of our model are: a nonstandard choice of the reference configuration, which allows us to cope with the continually evolving material structure; and a driving force and a kinetic law for accretion/ablation that involves the difference in chemical potential, strain energy and the radial stress. By combining these ingredients we arrive at an algebraic system which governs the stationary treadmilling state. We establish the conditions under which this system has a solution and we show that this solution is unique. 
Moreover, by an asymptotic analysis we show that for small beads the thickness of the solid is proportional to the radius of the support and is strongly affected by the stiffness of the solid, whereas for large beads the stiffness of the solid is essentially irrelevant, the thickness being proportional to a characteristic length that depends on the parameters that govern diffusion and accretion kinetics.

\end{abstract}


\section{Introduction.}\label{sec-1}

Surface growth, i.e. the accretion of a solid onto a surface, occurs in several contexts of physical, technological, and biological interest. One of the most common examples of surface growth is the solidification of water at the ice-water interface near the freezing temperature; other examples include technological processes such as chemical vapor deposition or, in biology, the growth of hard tissues like bones and teeth.

Although surface growth may be regarded as bulk growth concentrated on a surface \cite{DiCar2005Surface},  surface and bulk growth are in general treated in a different manner. When dealing with bulk growth, the reference configuration is fixed and the addition of particles to the body is accounted for by a tensor field, often referred to as the \emph{growth tensor} \cite{RodriHM1994JB}, whose value at a given point identifies the stress-free stance \cite{DiCarQ2002Growth} of a chunk of material in a small neighborhood of that point. When dealing with surface growth, on the other hand, it seems more natural to account for addition and removal of material by letting the boundary of the reference configuration evolve, as done in \cite{Skalak1997}. 

The choice of a reference configuration in bulk growth is rarely an issue: in fact, because of the extra degree of freedom brought in by the growth tensor, there is always a pair of a deformation and of a growth-tensor field that identifies a stress-free state.
For surface growth, on the contrary, it is not always obvious what is the most convenient choice of a reference configuration. In particular, it might be impossible to identify a stress-free state through the conventional notion of reference configuration in a three-dimensional reference space. From the kinematic standpoint, a resolution to this difficulty was provided by Skalak et al. \cite{Skalak1982}  in a seminal paper by introducing the time $\tau$ at which a material point is deposited on the growth surface. They label each particle of the body at time $t$ with four coordinates $(a_1, a_2,\tau, t)$, where $(a_1, a_2)$ are the coordinates of the point on the two-dimensional growth surface at which the material point was deposited.  It is this basic idea that we build upon in our developments.\color{black}

\comment{I removed the paragraph on Nash theorem. It made sense when it was accompanied by a discussion on the integrability of the deformation gradient. Now that this discussion has been removed, it may only confuse the reader.}

\color{red}

\color{black}Another important feature of surface growth is the dependence of the {accretion} rate on the local stress field.  Often, {accretion} happens at the outer surface of the body, with each new layer of material forming on top of the layer that was last formed. If each new layer is geometrically compatible with the previous one, {accretion} does not lead to a build-up of stress. On the other hand if {accretion} occurs on an interior surface of the body, and each new layer of material has to push away the previous layer, then this necessarily generates a residual stress in the body.\comment{I think this is not always true. In fact, the stress field is constant at treadmilling.}  \color{black}The stress field, by its turn, appears in the laws governing accretion rate through an Eshelbian-like coupling term \cite{AmbroG2007MMS,CiarlPM2013Mechanobiology,DiCarQ2002Growth,EpsteM2000Thermomechanics}. All these effects result into an intimate coupling between mechanics and growth.

\begin{figure}[h]
\begin{center}
{\includegraphics[width=0.6\textwidth]{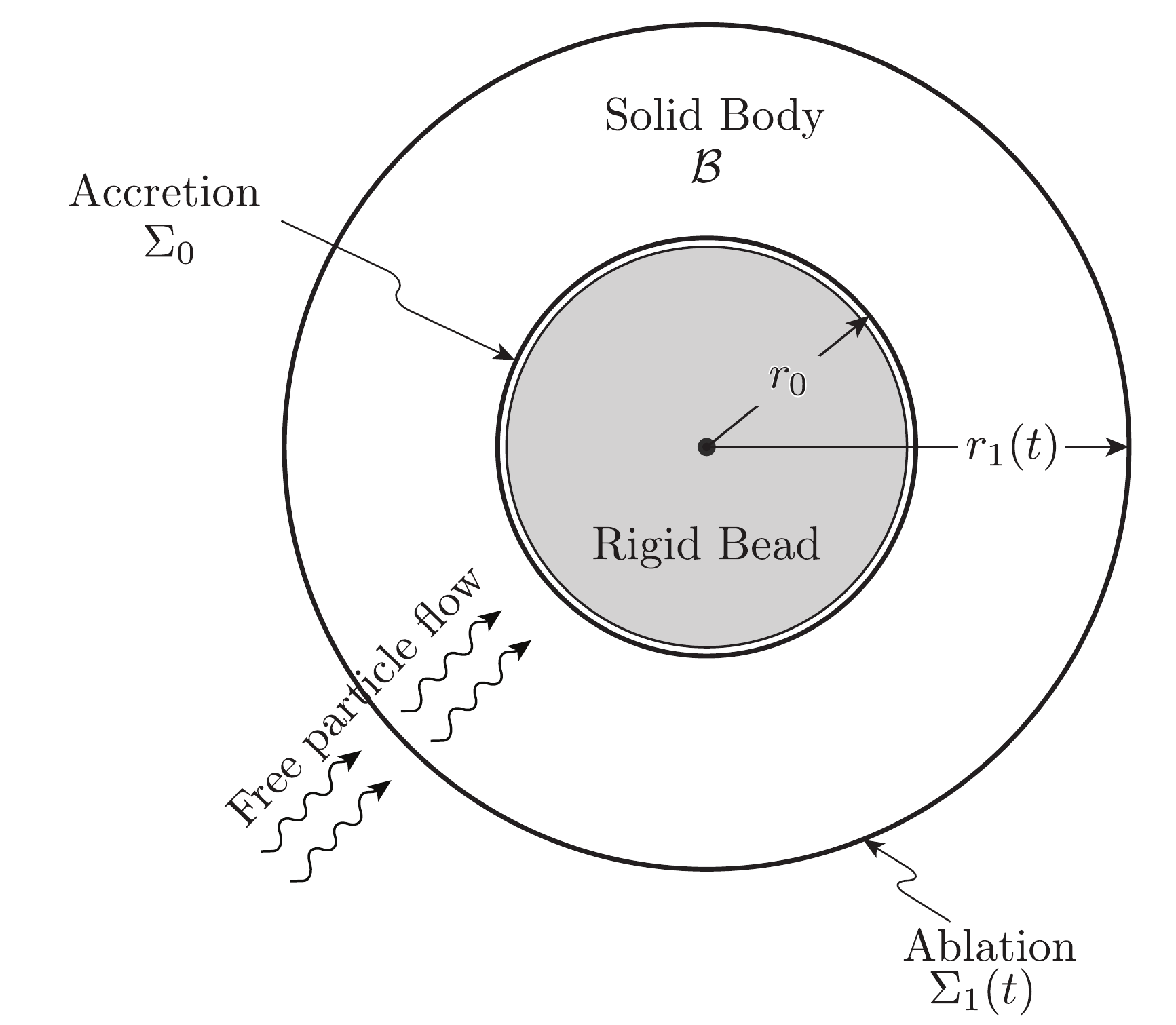}} 
\end{center}
\caption{\footnotesize Problem setting: An annular solid body ${\cal B}$, bounded by two concentric spherical surfaces $\Sigma_0$ and $\Sigma_1(t)$. The former represents the fixed support at which {accretion} occurs, the later is the (in general) time-dependent outer surface of the body.   A fluid containing ``free particles'' fills the entire region outside $\Sigma_0$, and therefore surrounds and permeates the solid.  The free particles diffuse through the solid to reach its inner surface $\Sigma_0$ where they attach to the solid. Free particles are continually being attached to the body at the inner surface and detached from the body at the outer surface.}
\label{F1}
\end{figure}

The specific problem we consider is described schematically in Figure \ref{F1}.  A note on terminology first: since we wish to refer to the individual units of material that combine to form a body, it is convenient to refer:to any such unit as a ``free particle''; to the body formed by the combination of many free particles as the ``solid''; to \color{black}the processes of adding and removing free particles from the solid as ``accretion'' and ``ablation''\color{blue}, \color{black}respectively.  In Figure \ref{F1} the solid body ${\cal B}$\comment{Just as a note to be kept commented in the source file: A reviewer may object that $\mathcal B$ is \textbf{not} the body, but the region in space occupied by the body. Also, one may argue that the body, regarded as a collection of particles (each identified by a unique label), is instead $\mathcal M\color{blue}(t)\color{black}$. This is in fact the point of view of Truesdell and Noll.} occupies the region between the two spherical surfaces $\Sigma_0$ and $\Sigma_1(t)$. The inner surface $\Sigma_0$ is the boundary of a rigid bead on whose surface accretion occurs. The bead is surrounded by a fluid in which the free particles are dispersed, the fluid occupying the entire region outside $\Sigma_0$, wherefore it both surrounds and permeates the solid. The free particles diffuse towards the bead surface due to a gradient in chemical potential. When they arrive at $\Sigma_0$ they attach  to ${\cal B}$.  At the same time, it turns out that it is energetically favorable for the solid to shed free particles at its outer surface $\Sigma_1(t)$, wherefore ablation occurs simultaneously at $\Sigma_1(t)$.  Thus free particles are continually being attached and detached from the body, the former\comment{The previous clause has only one subject, one complement, and no gerund. Thus, the only pair of nouns that this istance of ``former'' and ``latter'' may refer to is ``free particles'' and body'', respectively; then I do not understand the meaning of the sentence.} at the inner surface and the latter at the outer surface. If the rate of accretion is greater than the rate of ablation, the body grows and $\Sigma_1(t)$ moves outwards. The evolving stress and deformation fields within the solid are governed by a mechanics problem. The flow of free particles is governed by a diffusion problem. And these two problems are coupled at the two surfaces of the body through both the conservation of free particles and the kinetics of accretion \color{blue}(when there is no ambiguity, as in the preceding sentence, we shall use the term accretion to refer to both accretion at the inner surface and ablation at the outer surface).\comment{This sentence was in a footnote. I had it put within parentheses. Is that okay?} \color{black}We study the steady ``treadmilling'' problem where the accretion rate balances the ablation rate so that the outer radius of the body is in fact  time-independent (even though free particles continue to attach and detach at the two surfaces).

Although our aim is to illustrate certain ideas in continuum mechanics, the problem we consider was  inspired by an experiment  described by Noireaux et al. \cite{Noir2000} (see also \cite{CamerFVT1999Motility}). In  this experiment, a polymeric gel is grown on  a spherical bead by immersing it in a solution containing actin, a protein which can polymerize and form filaments entangled and cross-linked into an elastic network \cite{Theri2000T}. Previous chemical treatment of the  bead's surface ensures that actin polymerization --- and hence accretion of the network --- takes place on that surface; moreover, the permeability of the network to the surrounding solution makes it possible for the actin units in the solvent to diffuse towards the bead's surface, where they polymerize and attach to the network. The results reported in  \cite{Noir2000} show the existence of a treadmilling state

\color{black}There \color{black}is a vast and rapidly growing body of literature on the mechanics of growth which we shall not attempt to review here. The reader is referred to, for example, the review papers by Ambrosi et al.\cite{ArrudaKuhlGarikipati2011}, Garikipati \cite{garikipati}, Jones and Chapman \cite{jones}, Kuhl \cite{Kuhl2014}, Menzel and Kuhl \cite{kuhl2012}, and Taber \cite{Taber1995}, and to the book by Epstein \cite{epstein}. Examples with residual stresses induced in an elastic solid by volumetric growth in spherical symmetry may be found in \cite{AmbroM2002JMPS}, \cite{BenAG2005JMPS}, and \cite{MoultG2011JELAS}. An example involving surface growth is discussed in \cite{CiarlPM2013Mechanobiology}.
Concerning the aforementioned experiments, by which our problem was inspired,
Noireaux et al. \cite{Noir2000} examined this problem using linear elasticity.  Dafalias and Pitouras \cite{dafalias2009stress}
examined the mechanics aspects of this problem using particular finite elasticity constitutive models; see also \cite{dafalias2008stress, durban2015solid}.  A more recent study by Cohen et al. \cite{Cohen2014} considered the effect of dampening (resulting from the interaction between the solid matrix and the solvent flow) and investigated the time-dependent evolution of the system leading to the treadmilling regime.

The principal contributions of the present paper \color{blue}consist in\color{black}: the introduction of the notion of a four-dimensional reference space in characterizing surface growth; a finite deformation analysis of the mechanical problem for an arbitrary isotropic elastic body; the coupling of the chemical problem to the finite deformation mechanics problem; showing
that  it is the build-up of strain energy, not stress, that causes the ablation rate at the outer surface to increase as the body grows; \color{black} and the development of a {thermodynamically consistent} notion of the driving force for {accretion} that explicitly 
\color{blue}accounts for
\color{black}energy, stress and 
\color{blue}chemical potential.

\color{black}In Section \ref{sec-2} we formulate and solve the finite deformation mechanics problem for the elastic solid ${\cal B}$.  This problem is coupled to the free particle diffusion problem in two ways.  One is by the conservation of mass as the free particles are attached and detached from the solid body. This is addressed in Section \ref{sec-3}.  Then in Section \ref{sec-3MIT} we model the steady diffusion of free particles. The preceding effects are further coupled through the kinetics of {accretion}. This is formulated in Section  \ref{sec-4} where, in particular, we develop the {notion of the} driving force for {accretion}.   {We assume that the deviations from thermodynamical equilibrium are small, and take a linear kinetic relation between driving force and accretion rate.} Finally in Section \ref{sec-5} we study the steady response of the system, establishing precise conditions under which this system can have a treadmilling solution, and then examining in more detail the thickness of the body and the {accretion} rate in the limiting cases of a small bead 
\color{blue2}(the stress--limited regime) 
\color{black}and a large bead 
\color{blue2}(the diffusion--limited regime)\color{black}. The implications of the results are discussed in Section \ref{sec-8} and we end with a brief summary.


\section{Mechanics of the solid body.}   \label{sec-2}

\noindent \color{blue}In this section we first propose a notion of reference configuration, deformation and strain for the solid body. Then, we determine the stress field and the strain energy density within the \emph{body manifold} $\mathcal B(t)$, namely, the region occupied by the body at time $t$ in the physical space. The ingredients of our construction are: a three-dimensional \emph{material manifold} $\mathcal M(t)$ whose elements identify the 
material points
that comprise the body at time $t$, the manifold $\mathcal M(t)$ being immersed in a four-dimensional reference space; a \emph{placement map} $\bm\chi(\cdot,t)$ which assigns to the typical particle $\mathbf X\in\mathcal M(t)$ the position $\bm x=\bm\chi(\mathbf X,t)$ that the particle occupies at time $t$; a constitutive equation relating the deformation gradient and the stress. \color{black}

\bigskip
\noindent\textbf{The physical space.} 
We shall identify with $\mathbb R^3$ the physical space where the 
\color{blue}motion\comment{I think Tal's concern whas that ``motion of the body'' may suggest motion as a whole without deformation.}
\color{black}takes place. Under the present circumstances, it is natural to label points in the physical space by spherical coordinates $(r,\theta,\phi)$ such that a typical point is represented by 
\begin{equation}
 \bm x=\bm x(r,\theta,\phi):=\left(r\cos\theta\cos\phi,r\sin\theta\cos\phi,r\sin\phi\right). \label{eq:1}
\end{equation}
 {The body occupies the region between two concentric spherical surfaces $\Sigma_0$ and $\Sigma_1$,}  so that for $r=r_0$ and $r=r_1$, $\bm x$ lies on, respectively, the inner boundary $\Sigma_0$ and the outer boundary $\Sigma_1$. \color{blue}For later use, we introduce the following orthonormal basis\color{black}
\begin{equation}
\begin{aligned}\label{eq:30}
 & \mathbf e_r:=\frac{\partial\bm x}{\partial r},\qquad  \mathbf e_\theta:=\frac{1}{r}\frac{\partial\bm x}{\partial \theta},\qquad \mathbf e_\phi:=\frac{1}{r}\frac{\partial\bm x}{\partial\phi}.
\end{aligned}
\end{equation}


\color{black}In order to unambiguously identify  individual particles during their motion in physical space, we now introduce  a \textit{material manifold} $ \mathcal{M}(t)$ within a reference space.
\bigskip


\noindent\textbf{The reference space.} Surface growth  involves two uncommon features that need some attention:  
$(i)$ the material manifold  $\mathcal{M}(t)$ does not constitute a fixed collection of material points 
and $(ii)$ as we shall see in what follows, the solid material is formed in physical space under stressed conditions and  the material accumulates residual stresses such that, even when 
\color{blue}the support is removed,
\color{black}the grown body is not stress-free.  The choice of the particular material manifold and reference space is made so as to model these effects as simply as possible.

Given that particles are sequentially added to the body at the inner surface  $\Sigma_0$, we choose, as material manifold, the Cartesian product
\begin{equation}\label{eq:1}
 \mathcal{M}(t)=\Sigma_0\times(Z_0(t),Z_1(t))
\end{equation}
between the constant spherical surface $\Sigma_0$ on which {accretion} occurs and an open, time--varying interval $(Z_0(t),Z_1(t))$. The set defined in \eqref{eq:1}  is a smooth submanifold of a four-dimensional \emph{reference space} $\mathbb R^4$:
\begin{equation}
\mathcal {M}(t)\subset\mathbb R^4.
\end{equation}
This material manifold is in fact a \color{blue2}cylindrical hypersurface \color{black}parallel to the axis $Z$,
\color{blue}each cross section being a copy of the surface $\Sigma_0$ on which accretion takes place. Although both $\mathcal B(t)$ and $\mathcal M(t)$ may depend on time, hereafter we shall omit such dependence when there is no risk of confusion.\color{black}

\begin{figure}[h]
\begin{center}\includegraphics[width=0.9\textwidth]{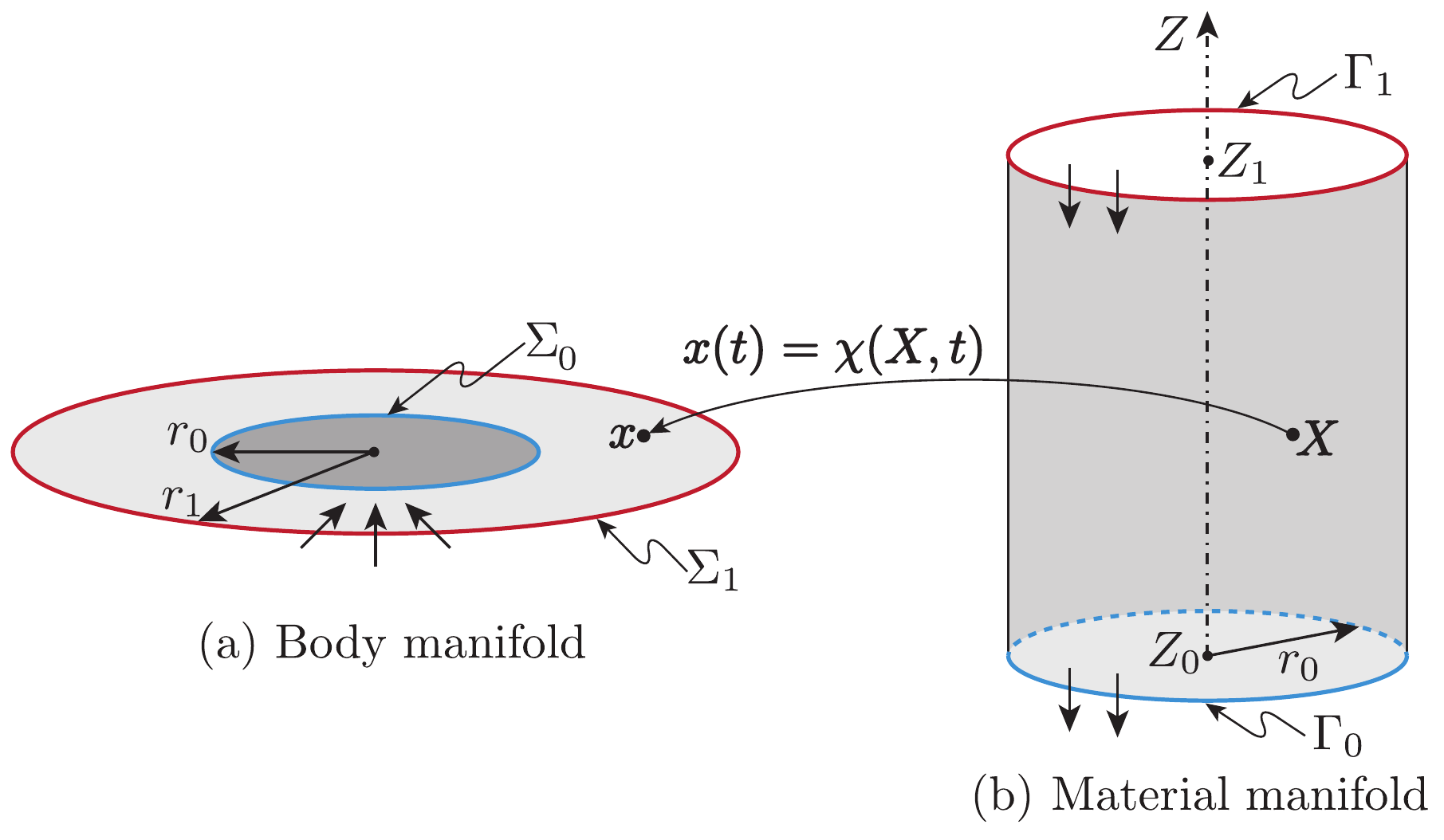} \end{center}
\caption{Schematic depiction of the body manifold $\mathcal B\color{blue}\color{black}$ and the material manifold $\mathcal M\color{blue}\color{black}$ at a given time $t$ in the case of a two dimensional physical space. The annular disk on the left-hand side \color{black}represents  the body manifold while the cylinder on the right-hand side \color{black}is the material manifold. The radius $r_0$ of the cylinder equals the inner radius of the annular region. The two ends of the cylinder, $\Gamma_0$ and $\Gamma_1$, can move in the vertical direction thus adding or removing material points to $\mathcal M$.}
\label{F2}
\end{figure}

In order to gain some intuitive understanding of the material manifold, 
\color{black}it is helpful to consider the analogous lower dimensional problem  where growth takes place on a circular ring in a two-dimensional space as illustrated in Figure \ref{F2}(a).  The \color{blue}body manifold \color{black}$\mathcal B(t)$ in physical space is therefore a circular annular disc with inner and outer radii $r_0$ and $r_1(t)$.  The material manifold $\mathcal {M}$ in reference space, defined by (\ref{eq:1}) and shown schematically in Figure \ref{F2}(b),  is a cylinder in three-dimensional space.  Its ends, $Z = Z_0(t)$ and $Z=Z_1(t)$ (or $\Gamma_0(t)$ and $\Gamma_1(t)$ in Figure \ref{F2}(b)), correspond to the respective boundaries $\Sigma_0$ and $\Sigma_1(t)$ in physical space. Since new material is continually being added at $\Sigma_0$ and removed at $\Sigma_1$, the ends of the cylinder $\mathcal {M}$ in the material manifold will be time dependent in general and the cylinder may change its length. As depicted by the arrows in Figure \ref{F2}(b), when material is being added at $\Gamma_0$, the boundary $\Gamma_0$ must 
\color{blue}translate parallel to the $Z$-axis along the negative direction so as to incorporate new
\color{black}material points into ${\cal M}$; similarly when material is being lost at $\Gamma_1$, the boundary $\Gamma_1$ must 
\color{blue}translate
\color{black}in the negative $Z$-direction 
\color{blue}so that material points are removed from ${\cal M}$.
\color{black}The rate at which material is added at $\Sigma_0$, as characterized by the motion of $\Gamma_0$ in reference space, is therefore given by $-2\pi r_0 \dot{Z}_0(t)$.  Similarly the rate at which material is removed at $\Sigma_1$ is  $-2\pi r_0 \dot{Z}_1(t)$. Thus, for example, the rate at which the incompressible body expands in physical space, $ 2\pi r_1 \dot{r}_1(t)$ must equal  $-2\pi r_0 \dot{Z}_0(t) + 2\pi r_0 \dot{Z}_1(t)$. We will encounter analogous expressions in our higher dimensional problem.  It is worth emphasizing that in physical space, even though the radius $r_0$ of the inner surface is constant, the radial velocity of the material points at $r=r_0$ does not vanish.  In fact \color{blue} the radial velocity \color{black}is  $-\dot{Z}_0$, as can be deduced from the fact that the rate of material addition at $r=r_0$ is $-2\pi r_0 \dot{Z}_0(t)$.

Returning to our (higher dimensional) setting, it is natural to describe the material \color{black}manifold through the parametric characterization
\begin{equation} \label{eq:5}
\mathbf X=\mathbf X(\Theta,\Phi,Z):=(r_0\cos\Theta\cos\Phi,r_0\sin\Theta\cos\Phi,r_0\sin\Phi,Z),
\end{equation}
where $0 \leq \Theta < 2 \pi, \, 0 \leq \Phi \leq \pi,  \, Z_0 \leq Z \leq Z_1$. 
\comment{I removed the  sentence: ``Observe that ${\bf X} \in {\mathbb R^4}$ though ${\bf x}\in {\mathbb R^3}$.'' Although our previous discussion mentions that $\bm x$ corresponds to $\mathbf X$, we have not introduced the deformation map. Thus, mentioning $\bm x$ at this point interrupts the flow and makes reading more difficult.}

\color{black}We interpret the coordinates $(\Theta,\Phi,Z)$ of the typical point $\mathbf X$ as follows:
\begin{itemize} 
\item the pair $(\Theta,\Phi)$ specifies where on $\Sigma_0$ the particle  $\mathbf X$ was added to the body; 

\comment{I could see many problems with the text that was there before. First: a ``sequence'' is not something that can be ``measured''. Second: directionality and sign are distinct concepts and should not be confused. Third: time can be negative as well (otherwise the concept of invariance with respect to time translation would be meaningless. Finally, and this is the most problematic point, the ordering induced by $Z$ is exactly the opposite of the chronological ordering. In fact, if $Z_a<Z_b$ then the particles labeled with $Z_a$ are added to the body later than the particles labeled with $Z_b$. Finally, I am not sure that ``chronological time'' and ``time'' differ in their meanings. I think we should refrain from referring to $Z$ as chronological time. }

\item \color{blue}The variable $Z$ identifies a section of the cylinder (a sphere of radius $r_0$) comprising all material points added to the body at the same instant. We denote the corresponding time by $t_0(Z)$.
\end{itemize}
\color{blue}In our construction the variable $Z$ shall specify the (reverse) order in which material points are added to the material manifold: if two material particles, say $\mathbf X_a$ and $\mathbf X_b$ belong to sections $Z_a$ and $Z_b$ with $Z_a<Z_b$, then  $\mathbf X_a$ has been added to the body after  $\mathbf X_b$. In particular the section $Z_0(t)$ contains all material points added to the body at time $t$.

\color{black}As mentioned in the introduction, our choice in labeling material points is inspired by the proposal set forth in \cite{Skalak1982} (see also \cite{Skalak1997}), where it was suggested that the collection of material points that comprise a growing body at time $t$ be labeled through a triplet of coordinates $(\theta_1,\theta_2,\tau)$, with $\tau\le t$ denoting the time at which a particular point was added to the body. \color{blue}In our case, however, the fourth coordinate $Z$ cannot be identified with the time  $t_0(Z)$ when a free particle attaches to the body. 

For later use, we introduce the normalized orthogonal  
\color{blue}basis \color{black}associated with the coordinate system $(\Theta,\Phi,Z)$ on $ \mathcal{{M}}$ 
\begin{equation}\label{eq:34a}
  \mathbf e^\Theta:={r_0}\frac{\partial \Theta}{\partial\bm X},\qquad \mathbf e^\Phi:={r_0}\frac{\partial \Phi}{\partial\bm X},\qquad\mathbf e^Z:=\frac{\partial Z}{\partial\bm X} .
\end{equation}

\smallskip

\noindent\textbf{The placement map and its gradient.} 
We may   specify the placement of 
\color{blue}the body
\color{black}in the physical space $\mathbb R^3$ through a \emph{placement map} $\bm\chi\color{blue}(\cdot,t)\color{black}:\mathcal {M}\color{blue}(t)\color{black}\to\mathbb R^3$ which assigns a place 
\begin{equation}\label{eq:8}
\bm x(t)=\bm\chi(\mathbf X,t)
\end{equation}
to the typical particle $\mathbf X=\mathbf X(\Phi,\Theta,Z)$. The time dependence is due to growth.  A natural characterization of the placement map from the material coordinates  $(Z,\Theta,\Phi)$ to the spatial coordinates $(r, \theta, \phi)$  is 
\begin{equation}\label{eq:8a}
\bm\chi(\mathbf X,t)={\bm x}(r(\Theta,\Phi,Z,t), \theta(\Theta,\Phi,Z,t),\phi(\Theta,\Phi,Z,t))
\end{equation}
where ${\bm x}(r, \theta,\phi)$, was defined in (\ref{eq:1}).

By making use of the chain rule we can now write the \emph{deformation gradient} of the mapping from $\mathcal {M}\color{blue}(t)\color{black}$ to $\mathcal {B}\color{blue}(t)\color{black}$ as
\begin{equation}\label{eq:F0}
\begin{aligned}
\mathbf F(t)=\frac{\partial\bm\chi}{\partial \mathbf X}&\stackrel{\eqref{eq:8a}}=\frac{\partial\bm x}{\partial \Theta}\otimes\frac{\partial \Theta}{\partial \mathbf X}+\frac{\partial\bm x}{\partial \Phi}\otimes\frac{\partial \Phi}{\partial \mathbf X}+\frac{\partial\bm x}{\partial Z}\otimes\frac{\partial Z}{\partial \mathbf X}, \\
&\stackrel{\eqref{eq:34a}}=\frac{1}{r_0}\frac{\partial\bm x}{\partial \Theta}\otimes {\mathbf e\color{black}}^\Theta+\frac{1}{r_0}\frac{\partial\bm x}{\partial \Phi}\otimes {\mathbf e\color{black}}^\Phi+\frac{\partial\bm x}{\partial Z}\otimes {\mathbf e\color{black}}^Z\color{black}.
\end{aligned}
\end{equation}

\noindent\textbf{Spherical symmetry.} Looking for spherically-symmetric solutions, we now  restrict attention to placement maps such that $r(\Theta,\Phi,Z,t)$ is independent of $\Theta$ and $\Phi$, and that, trivially, $\theta(\Theta,\Phi,Z,t)=\Theta$ and $\phi(\Theta,\Phi,Z,t)=\Phi$. Thus 
\begin{equation}\label{eq:25}
  r=r(Z,t), \qquad \theta=\Theta,\qquad \phi=\Phi.
\end{equation}
\color{blue}Put in another way, equation \eqref{eq:25} states that the particle that is added to the body at location $(r_0, \theta,\phi)$ at time $t_0(Z)$ is located, at time $t$, at $(r(Z,t), \theta, \phi)$. 
In  particular, the particles occupying the positions $(r_0, \theta, \phi)$ and $(r_1, \theta, \phi)$ belong to the sections $Z_0(t)$ and $Z_1(t)$ and so we additionally require that\color{black} 
\begin{equation} \label{eq:37}
  r(Z_0(t),t)=r_0~~~~\text{and}~~~~r(Z_1(t),t)=r_1(t).
\end{equation}
\comment{I removed the last sentence, whose purpose I could not understand. It is clear from the outset that $r_0$ is a constant. What we wrote would have given the impression that $Z_0$ is independent on time as well.}
On account of \eqref{eq:25}, 
\color{blue}the representation \eqref{eq:8a} of the placement map becomes\color{black}\comment{\eqref{eq:8a} is not the placement map. Is just a representation of the placement map.}
\begin{equation}\label{eq:2}
  \bm\chi(\mathbf X,t)=\bm x(r(Z,t),\Theta,\Phi)
\end{equation}
and the deformation gradient (\ref{eq:F0}) specializes to 
  \begin{equation}
\begin{aligned}
\mathbf F&(t)\stackrel{\eqref{eq:30}}=\lambda_\theta \left(\mathbf{e}_\theta\otimes \mathbf{e}^\Theta+\mathbf{e}_\phi\otimes \mathbf{e}^\Phi\right)+\lambda_r\mathbf{e}_{r}\otimes \mathbf{e}^Z \label{eq:F2}
\end{aligned}
\end{equation}
where we have set
\begin{equation} \label{eq:28}
  \lambda_r=\frac{\partial r}{ \partial Z},\qquad   \lambda_\theta=\frac r {r_0}.
\end{equation}
It is worth noting that in the classical elasticity problem ({\it without} growth) of the radial deformation of a spherical shell, if one identifies $Z$ with the radial coordinate of a particle in the undeformed configuration, then the principal stretches would be $  \lambda_r = {\partial r}/{ \partial Z}$ and $\lambda_\theta= r/Z$ the latter of which differs from \eqref{eq:28}$_2$.


\medskip

\noindent \textbf{Material response: incompressibility.}
We assume that the material comprising the solid body is \emph{incompressible} in the sense that
\begin{equation} \label{eq:33}
 {\rm det} \, \mathbf F=\lambda_r\lambda_\theta^2=1.
\end{equation}
   On taking  \eqref{eq:28}  into account, equation (\ref{eq:33})  translates into the differential equation\begin{equation}\label{eq:4}
  \frac{{\partial }r}{{\partial} Z}=\left(\frac {r_0}r\right)^2.
\end{equation}
Integrating \eqref{eq:4} and enforcing the first of \eqref{eq:37} yields
\begin{equation} \label{eq:7}
  {r^3(Z,t)}={r_0^3}+3r_0^2(Z-Z_0(t)).
\end{equation}
This equation gives, explicitly, the radial coordinate $r$ at time $t$ of a particle that was added to the solid body at 
\color{blue} time $t_0(Z)$. \color{black}Differentiating (\ref{eq:7}) with respect to $t$ at constant (``reference coordinate'') $Z$, gives the particle velocity field of the solid body:
\begin{equation} \label{eq:21MIT}
\mathbf{v}(r,t) = v(r,t)  \, \mathbf{e}_r = - \dot{Z}_0(t) \  \frac{r_0^2}{r^2} \,  \mathbf{e}_r.
\end{equation}
The divergence of this velocity field is readily seen to vanish in keeping with the requirement of incompressibility.
Observe also  that the speed of a material point on the growth surface is $v(r_0,t)= - {\dot{Z}}_0(t)$ and this does not vanish in general, even though the support is rigid and $r_0$ is independent of time. This is a consequence of the growth that occurs at $r=r_0$.

Observe from (\ref{eq:7}) and (\ref{eq:37})$_{2}$ that
$ r_1^3(t)={r_0^3}+3r_0^2(Z_1(t)-Z_0(t))$.
\color{blue2}This relation, 
when differentiated with respect to time, \color{black}leads to
\begin{equation} \label{eq:111-MIT}
4 \pi  r_1^2(t) \dot{r}_1(t) =  \big[- 4 \pi r_0^2 \dot{Z}_0(t)\big] -  \big[ - 4 \pi r_0^2 \dot{Z}_1(t) \big].
 \end{equation}
The left hand side of this equation characterizes the rate at which the volume of the body increases. The two terms on the right hand side represent the rates at which material is added to the body at $\Sigma_0$ and removed at $\Sigma_1$; this is precisely the higher dimensional counterpart of the equation presented in the paragraph above (\ref{eq:5}) in our\comment{I removed ``intuitive''.}
discussion of the lower dimensional case.  The importance of (\ref{eq:111-MIT}) is that we will encounter the two terms on its right hand side when we model the flux of free particles and their kinetics during addition to, and removal from, 
\color{blue}the body.\color{black}


\medskip

\noindent \textbf{Material response: energy and stress.}  Assume now that the incompressible solid can be modeled as an isotropic elastic material. \color{black}As such, it can be characterized through a {referential strain energy function} $\widetilde{W}(\mathbf F) = \widehat{W}(\lambda_1, \lambda_2, \lambda_3)$ where the 
\color{blue}symbols $\lambda_i$ denote
\color{black}the principal stretches\color{black};\comment{Readers may ask how these stretches are defined. In fact, I am not sure how the polar decomposition theorem would apply to $\mathbf F$.}  
\color{blue}We assume that $\widehat{W}$ and its first derivatives vanish\comment{It is quite dangerous to speak about unstretched configuration for the body. This configuration exists only in a four-dimensional reference space, and it is not clear what would the meaning of the energy in that setting. I would just say that the energy vanishes when the stretches are all equal to one.}
\color{blue}for 
\color{black}$\lambda_1=\lambda_2=\lambda_3=1$. \color{black}The principal Cauchy stress components are then given by the constitutive
\color{blue}equation\color{black}
\begin{equation} \label{eq:14}
\sigma_k = \lambda_k \widehat{W}_k  - p, \qquad  \widehat{W}_k = {\partial \widehat{W}}/{\partial \lambda_k},  \qquad k = 1,2,3,
\end{equation}
where the pressure $p$ is constitutively indeterminate. 

Since we are dealing with isochoric equi-biaxial deformations characterized by
$$\lambda_1 = \lambda^{-2}, \qquad \lambda_2 = \lambda, \qquad \lambda_3 = \lambda,
$$ 
it is convenient to introduce a \emph{reduced strain energy}  ${W}(\lambda)$ \color{blue}defined \color{black}by
\begin{equation}\label{20150114.rhotel11}
{W}(\lambda) := \widehat{W}(\lambda^{-2}, \lambda, \lambda), \qquad \lambda >  0. 
\end{equation}
\color{blue}Our previous assumptions on $\widehat W$ guarantee that ${W}(1) =  {W}'(1)  = 0$. 

To examine the properties of the reduced strain energy function, consider the principal Cauchy stress components in isochoric equi-biaxial deformations under plane stress conditions, i.e. when $\sigma_1 = 0$ and  $\lambda_1 = \lambda^{-2}, \lambda_2 = \lambda, \lambda_3 = \lambda$. These are readily calculated from (\ref{eq:14}) and  (\ref{20150114.rhotel11}) to be
\begin{equation}
\sigma_1 = 0, \qquad \sigma_2 = \sigma_3 := \sigma = \frac{\lambda}{2} W'(\lambda).
\end{equation}
It is natural to require the equi-biaxial stress $\sigma$ to be tensile for $\lambda > 1$ and compressive for $0 < \lambda < 1$. Thus we shall endow $W$ with the properties
\begin{equation} \label{20150617.v2777}
W'(\lambda) > 0  \ \   {\rm for} \ \lambda > 1, \qquad W'(\lambda) < 0  \ \  {\rm for} \ 0 < \lambda < 1,
\end{equation}
which in particular imply that $W(\lambda)>0$ for $\lambda\neq 0$. \color{blue}In addition, we shall require that the reduced energy blows up under extreme elastic strains:\color{black}
\begin{equation} \label{eq:RA2}
W(\lambda) \to \infty \qquad {\rm as} \ \ \  \lambda \to \infty.
\end{equation}
\color{blue}This assumption, as we shall see later, ensures the existence of a treadmilling state.\color{black}

When applied to the problem analyzed in the present paper, we make the identification
\begin{equation}
\lambda=\lambda_\theta=\lambda_\phi,
\end{equation}
so that the Cauchy stress is given by 
\begin{equation}\label{eq:12}
 \boldsymbol{\sigma}= \sigma_r \mathbf e_r\otimes\mathbf e_r+\sigma_\theta  \left(\mathbf{e}_\theta\otimes \mathbf{e}_\theta+\mathbf{e}_\phi\otimes \mathbf{e}_\phi\right),
\end{equation}
where, in accordance with \eqref{eq:14}, the radial and the circumferential stress are given by, respectively, \color{black}
\begin{equation} 
\sigma_r = \lambda_r\widehat{W}_1(\lambda_r,\lambda_\theta,\lambda_\theta)  - p,\quad\text{and} \quad \sigma_\theta=\lambda_\theta\widehat{W}_2(\lambda_r,\lambda_\theta,\lambda_\theta)  - p.
\end{equation}
\medskip


\noindent\textbf{Equilibrium.} 
\color{blue}On taking into account spherical symmetry, and on recalling \eqref{eq:12}, we see that
\color{black}the equilibrium equation ${\rm div} \, \boldsymbol{\sigma}=\mathbf 0$ has only one non-trivial scalar consequence:
\begin{equation}\label{eq:srstra}
  \begin{aligned}
    \frac{{\partial} \sigma_r }{{\partial}r}+\frac 2 r\left(\sigma_r-\sigma_\theta\right)=0.
  \end{aligned}
\end{equation}


\color{blue}\noindent Now, by making use of the constitutive equation (\ref{eq:14}) and of the definition of $W(\lambda)$ given in \eqref{20150114.rhotel11}, and by noting from (\ref{eq:28}) and 
\eqref{eq:33} that  $\lambda_r = r^2_0/r^2$ and $\lambda_\theta = r/r_0$, we readily see that
\begin{equation} \label{eq:srst}
\begin{array}{lll}
\sigma_r - \sigma_\theta &=& \lambda_r \widehat{W}_1(\lambda_r, \lambda_\theta, \lambda_\theta) - 
\lambda_\theta \widehat{W}_2(\lambda_r, \lambda_\theta, \lambda_\theta) =\\[2ex]

&=& \displaystyle  r^2_0/r^2 \, \widehat{W}_1(r^2_0/r^2,  r/r_0,  r/r_0) -  r/r_0 \, \widehat{W}_2(r^2_0/r^2,  r/r_0,  r/r_0) = \\[2ex]

&=& \displaystyle - \frac 12 \, \frac{r}{r_0}\,  {W}'(r/r_0).
\end{array}
\end{equation}
\color{black}Therefore\color{blue}, the equilibrium  equation \color{black}(\ref{eq:srstra}) takes the\comment{Removed ``explicit''.} 
form
\begin{equation} \label{eq:10}  
\frac{\partial \sigma_r}{\partial r} = \frac{1}{r_0} {W}'(r/r_0). 
\end{equation}
\color{blue}On \color{black}integrating \eqref{eq:10} with respect to $r$ and 
\color{blue}on requiring that $\sigma_r$ vanishes\comment{Removed ``at all times''} on the outer surface $\Sigma_1$ (i.e., for $r=r_1$) we obtain \color{black}
\color{blue}an 
\color{black}explicit expression for the radial stress field in the\comment{Removed the adjective ``solid''} 
body
\begin{equation}  \label{eq:20}
{\sigma_{r}(r,t) = {W}(r/r_0) -  {W}(r_1(t)/r_0) }.
\end{equation}
The circumferential stress $\sigma_\theta(r,t)$ can now be determined through (\ref{eq:srst}).
\color{blue}In particular, by making use of (\ref{eq:20}) and (\ref{eq:srst}), and by recalling that  ${W}(1) = {W}'(1)  = 0$, we can compute the Cauchy stress components at the inner surface $r=r_0$:\color{black}
\begin{equation}\label{eq:21}
\sigma_{r}(r_0,t) = \sigma_{\theta}(r_0,t) = -  {W}(r_1(t)/r_0).
\end{equation}
Observe from (\ref{eq:21}) that, at the inner boundary $\Sigma_0$,
\begin{itemize}
\item[--] the\comment{Removed ``state of''.}
stress is hydrostatic;
\item[--]  the radial and circumferential stresses are both compressive;
\item[--]  the 
stress is, in general, time dependent due to its dependence on the time varying outer radius $r_1(t)$; 
\item[--]  and, as one might expect, the magnitudes of the stress components\comment{Removed ``here''} 
increase with the outer radius $r_1$.
\end{itemize}
Thus material is added to the body (in physical space) at a hydrostatically stressed state. In the steady (treadmilling) regime the radius $r_1$ is constant, and so the stress will also remain constant. 

As for the circumferential stress at the outer surface, since $\sigma_r(r_1(t),t)=0$, we have 
\begin{equation}\label{eq:18}
\sigma_\theta(r_1(t),t)=\frac 12 \frac {r_1(t)}{r_0}W'(r_1(t)/r_0)
\end{equation}
by \eqref{eq:srst}.  The right-hand side of \eqref{eq:18} is positive in view of our stipulation that $W'(\lambda)$ be positive for $\lambda>1$. Therefore the circumferential stress is tensile at the outer surface while, as noted previously, it is compressive at the inner surface.
 
Returning  to equations  (\ref{eq:28}) and (\ref{eq:33}) we observe that on the growth surface $r=r_0$ incompressibility implies $\lambda_r=\lambda_\theta=1$. However this does not mean that the body is unstressed at $r=r_0$ as is seen from (\ref{eq:21}).

 In summary, the radial velocity field $v(r,t)$ in the solid is given by (\ref{eq:21MIT}) and  the radial stress field $\sigma_r(r,t)$  is given by (\ref{eq:20}).
These expressions involve the as yet unknown outer radius $r_1(t)$ and the velocity ${\dot{Z}}_0(t)$ of the material points at the inner surface $r=r_0$.
 These functions depend on the accretion rate, and this in turn depends on both the local concentration of free particles, and the local stress and energy. This coupled chemo-mechanical process will be treated in the next sections.

\begin{figure}\label{fig:2000}
\begin{center}
\centerline{\includegraphics[scale=.5]{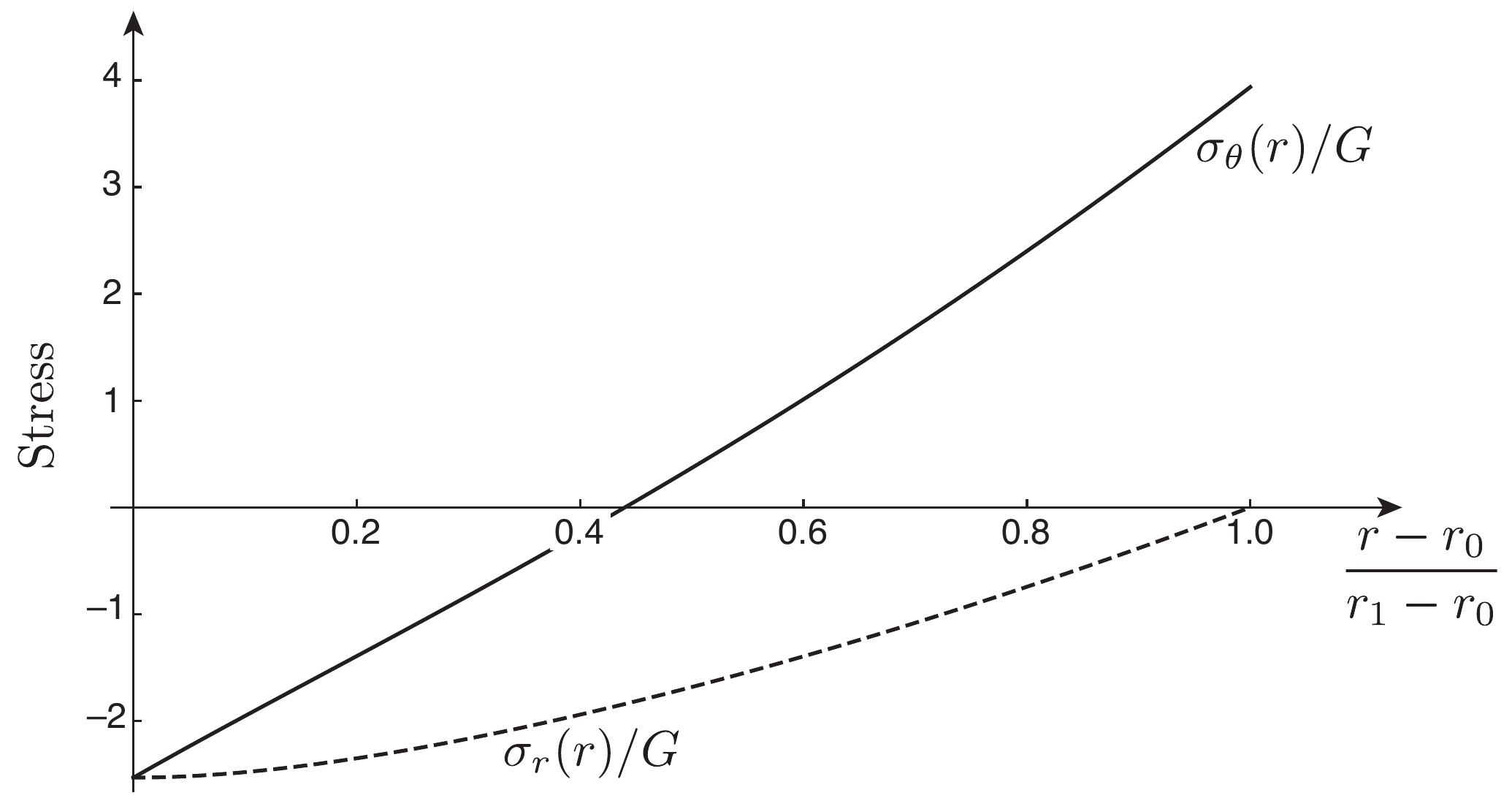}}
\end{center}
\caption{Renormalized radial and circumferential stresses for a neo-Hookean stored energy. The radial stress is compressive up to the outer surface, where it vanishes. The circumferential stress is compressive at the innter surface and tensile at the outer surface.}
\end{figure}

{{The numerical calculations in later sections will be carried out for}} a neo-Hookean material characterized by the strain energy function  
$$
\widehat W(\lambda_1,\lambda_2,\lambda_3) = \frac G2 \Big( \lambda_1^2 + \lambda_2^2 + \lambda_3^2 - 3\Big)
$$ 
with $G>0$ the shear modulus. In this case, the reduced energy $W(\lambda)$ defined in \eqref{20150114.rhotel11} is
\begin{equation}\label{eq:32}
  W(\lambda)=\frac G 2\left( \lambda^{-4} +2\lambda^2-3\right),
\end{equation}
and so the radial stress is given by
\begin{equation*}
{  \frac{\sigma_{r}(r, t)}{G}=\frac 1 2 \left[\Big(\frac {r_{0}}{r}\Big)^4-\Big(\frac  
{r_{0}} {r_{1}(t)}\Big)^4\right]+\Big(\frac {r}{r_{1}(t)} \Big)^2-\Big(\frac 
{r_{1}(t)} {r_{0}}\Big)^2\color{blue}.\color{black}}
\end{equation*}
\color{blue}The circumferential stress can now be recovered from the expression for the stress difference obtained in \eqref{eq:srst}, which for $W(\lambda)$ as in \eqref{eq:32} yields\color{black}
\[
{\frac{\sigma_{r}(r,t)-\sigma_{\theta}(r,t)}G=
\lambda_r^2(r,t)-\lambda_r^{-1}(r,t)=
\Big(\frac {r_{0}}r\Big)^4-\Big(\frac{r}{r_{0}}\Big)^2 .}
\]
Plots of the radial and circumferential stresses are shown in Figure \ref{fig:2000}. \color{blue}These plots show that the radial stress is everywhere compressive, whereas the circumferential stress is compressive near the inner surface and tensile in the proximity of the outer surface.\color{black}

\color{black}


\section{{Conservation of mass.} }\label{sec-3}

As mentioned previously in Section \ref{sec-1}, the elastic solid and surrounding fluid, exchange free particles at the surfaces $\Sigma_0$ and $\Sigma_1$. 
 In the presence of spherical symmetry, the free particles diffuse in the radially inward direction (due to a chemical potential gradient as will be discussed in the next section). When the diffusing free particles reach the inner surface $\Sigma_0$ they are removed from the fluid and attached to the solid. Thus the incoming free particle flux at $r=r_0$ is {balanced by} the rate of accretion of the solid. Similarly at the outer surface $\Sigma_1$, the inward flow of free particles from the region outside the solid body ($r > r_1$) crosses this surface and continues as an inward radial flow of 
\color{blue}free particles\color{black}.\comment{Hereafter I found several instances of the word ``diffusant''. I replaced them with ``free particles''.}  \color{black}However\color{blue}, \color{black}at $r = r_1$, free particles are being removed from the solid and added back into the fluid. Thus the free particle flux is not continuous at $r= r_1$ and its value will change discontinuously (jump) by an amount {equal} to the rate of ablation of the solid.

 We now formalize the preceding description by making the following additional modeling choices:
\begin{itemize}
\item [(A1)] the supply of mass added or removed from ${\mathcal M}$ is provided by a diffusant dispersed in an incompressible fluid that occupies the entire space outside of $\Sigma_0$;
\item [(A2)] particles can be added to and removed from the material manifold ${\mathcal M}$ only at its boundary $\partial{\mathcal M}=\Gamma_0\cup\Gamma_1$. 
\end{itemize}\medskip
On account of (A1), we introduce a spatial scalar field $\varrho$ and a spatial vector field $\mathbf h$, defined everywhere outside $\Sigma_0$, and representing the \emph{density} and \emph{flux} of diffusant\color{blue}, \color{black}respectively.  
Granted spherical symmetry, we may assume the diffusant flux to be radial,
\begin{equation}\label{eq:40}
 \mathbf h=h(r,t)\mathbf e_r,
\end{equation}
where $- h(r,t)$ is the mass of 
\color{blue}free particles
\color{black}that crosses a unit area of a spherical surface of radius $r$, in the radially {\it inward} direction,  in unit time.

In order to link this 
\color{blue}flux of free particles
\color{black}to the  {accretion} \color{blue}rate \color{black}we make the following hypothesis:
\begin{itemize}
\item [(A3)]  in order to build-up a unit volume of ${\mathcal M}$, the mass \color{black}of 
\color{blue}free particles 
\color{black}that must be converted into solid particles is a positive {\it constant} $\varrho_R$. 
\end{itemize}

We first enforce a balance between the flux of free particles diffusing through the fluid and arriving at $\Sigma_0$ and the rate at which they are added to the solid through {accretion}.  The mass of 
\color{blue}free particles
\color{black}arriving at $\Sigma_0$ per unit time  is $-4\pi r_0^2h(r_0,t)$.
Since there 
\color{blue}are no free particles
\color{black}in the interior of $\Sigma_0$, this serves as the sole free particle supply to the material manifold at $\Gamma_0$.   
\color{blue}We next consider the material manifold.
\color{black}Since  the outward normal velocity of its boundary $\Gamma_0$ is $-\dot{Z}_0(t)$ (see Figure \ref{F2}(b)), the rate at which free particles are added to ${\mathcal M}$ at $\Gamma_0$ is $4\pi r_0^2(- \dot{Z}_0) \varrho_R$; see discussion surrounding \eqref{eq:111-MIT}.   Mass conservation requires this 
\color{blue}rate 
\color{black}to equal the rate at which diffusant is lost from the fluid at $\Sigma_0$ and so we must have $-4\pi r_0^2h(r_0,t) = 4\pi r_0^2 (- \dot{Z}_0)\varrho_R$. This leads to
\begin{equation} \label{eq:47}
 h(r_0,t) = - {\varrho_R}V_0(t)
\end{equation}
where we have introduced the \emph{{accretion} \color{blue}{rate}}:
\begin{equation}\label{eq:39}
  V_0:=-\dot Z_0,
\end{equation}
at the inner surface.\color{black}

Second we consider the corresponding issue at the outer surface. 
The radially inward flux of 
\color{blue}free particles
\color{black}increases discontinuously from $- h(r_1+,t)$ to $-h(r_1-,t)$ as it crosses $\Sigma_1$.
This increase
 is due to the 
\color{blue}free-particle
\color{black}supply resulting from ablation of the material manifold at $\Gamma_1$. 
Now consider the material manifold.  Since the outward normal velocity of its boundary $\Gamma_1$ is $\dot{Z}_1(t)$  (see Figure \ref{F2}(b)), the rate at which material is removed from ${\mathcal M}$ at $\Gamma_1$ is $-4\pi r_0^2 \dot{Z}_1 \varrho_R$; see discussion surrounding \eqref{eq:111-MIT}.    Mass conservation requires   $- 4\pi r_0^2 \dot{Z}_1 \varrho_R  = [- 4\pi r_1^2 h(r_1-,t)] - [- 4\pi r_1^2 h(r_1+,t)]$
which leads to
\begin{equation} \label{eq:44}
h(r_1+,t)  - h(r_1-,t)  = - \left({r_0}/{r_1}\right)^2{\varrho_R}V_1(t),
\end{equation}
where we have introduced 
\begin{equation} \label{eq:1MITv2}
  V_1:=\dot Z_1.
\end{equation}

Finally, the following equation will be useful in what follows and so we record it here: observe from \eqref{eq:111-MIT}, \eqref{eq:39} and \eqref{eq:1MITv2} that
\begin{equation}  \label{eq:v2-114MIT}
r_1^2 \dot{r}_1 = r_0^2 (V_1 + V_0),
\end{equation}
where $\dot{r}_1(t)$ is the rate of increase of the radius of the outer surface $\Sigma_1$ of the solid.

The pair of equations \eqref{eq:47} and \eqref{eq:44} characterize the conservation of {free particle mass} during the accretive processes at $r=r_0$ and $r=r_1$ and couples the mechanics problem to the diffusion problem. They involve the {diffusant} fluxes $h(r_0,t)$ and $h(r_1\pm, t)$, and the material manifold boundary velocities $V_0$ and $V_1$.  In order to proceed further we need additional information on the fluxes.  Since the diffusant flux is driven by a gradient of the chemical potential, we now consider the role of the chemical potential in the diffusive process.

\section{Diffusion of free particles.}\label{sec-3MIT}

The conservation of free particles requires that the free particle density $\varrho$ and the free particle flux ${\mathbf h}$ obey the balance equation $\dot\varrho+{\rm div} \, \mathbf h=0$ 
away from $\partial\mathcal B$.  In the simplest models of diffusion, the flux is further assumed to obey Fick's law,
$\mathbf h = - M \nabla\mu$, with the chemical potential $\mu$ being related constitutively to $\varrho$ by an equation of state of the form $\mu = \varphi'(\varrho)$ where $\varphi$ is a free energy function. 
We shall simplify the analysis here by
\begin{itemize}
\item [(A4)] limiting attention from hereon to steady state evolution processes where all spatial fields are independent of time.
\end{itemize}
Since $\mathbf h = h(r) {\bm e}_r$ under steady spherically symmetric conditions,  the mass balance equation  simplifies to  
\be \label{eq:11ra}
{\rm div} \, \mathbf h =  \frac{1}{r^2} \frac{\partial}{\partial r} (r^2h(r)) = 0.
\ee
Integrating this and enforcing 
\color{blue}mass balance through the requirements
\color{black}\eqref{eq:47} and  \eqref{eq:44} leads to 
\be 
h(r) = \left\{
\ba{lll}
\displaystyle - \varrho_R V_0 \, \frac{r_0^2}{r^2}, \qquad \qquad & \displaystyle r_0 < r < r_1,\\[2ex]

\displaystyle - \varrho_R (V_0+V_1) \, \frac{r_0^2}{r^2}, \qquad \qquad & \displaystyle  r > r_1.\\
\ea
\right.
\label{eq-20151128-1}
\ee

Next we take the flux of free particles to obey Fick's law \color{black} and we allow\color{black}
\begin{itemize}
\item [(A5)] the scalar diffusion mobility $M$ in Fick's law to have different (constant) values, $M^-$ and $M^+$, in the regions inside and outside the solid respectively.
\end{itemize}
Thus we write $\mathbf h = - M \nabla\mu$ in the explicit form\comment{I switched to $\partial$ for consistency with our previous notation}
\begin{equation}\label{eq:9}
h = - M  \frac{\color{blue}\partial\color{black}\mu}{\color{blue}\partial\color{black}r},\qquad
  M(r) = \left\{
  \begin{array}{lll}
M^- \, (>0), \quad r_0 < r < r_1,\\[2ex]
  
M^+ \, (>0), \quad r > r_1,\\
\end{array}
\right.
\end{equation}
where we have used the fact that under steady spherically symmetric conditions $\mu = \mu(r)$.
We shall assume that
\begin{itemize}
\item [(A6)] $\mu(r) \to \mu_\infty$ as  $r \to \infty$\color{blue}, with $\mu_\infty$ a prescribed remote value of the chemical potential.\color{black} 

\end{itemize}

Substituting \eqref{eq-20151128-1} into \eqref{eq:9}, integrating, \color{blue}and enforcing (A6) \color{black}leads to
\be\label{eq-20151128-2}
\mu(r) =
\left\{
\ba{lll}
\displaystyle \mu_0 + \frac{\varrho_Rr_0V_0}{M^-} \left( 1 - \frac{r_0}{r}\right), \qquad & \displaystyle r_0 \leq r<r_1,\\[3ex]

\displaystyle \mu_\infty  - \frac{\varrho_R(V_0+V_1)}{M^+}  \frac{r_0^2}{r}, \qquad & \displaystyle r \geq r_1,\\
\ea
\right.
\ee
\color{blue}where $\mu_0=\mu(r_0)$,\comment{Here inverted order of $\mu_0$ and $\mu(r_0)$ for consistency with formulas below} the chemical potential at the inner surface $\Sigma_0$, is an unknown to be determined later\color{black}.\comment{Newline inserted here.} 

Finally, requiring the chemical potential  to be continuous at $r=r_1$ and letting $\mu_1 = \mu(r_1)$ denote its value there, we obtain the following pair of equations from \eqref{eq-20151128-2} :
\begin{equation} \label{eq:52}
\begin{aligned}
  {\varrho_R} V_{0}= M^-  \, \frac{\mu_1-\mu_0}{r_1-r_0}\frac {r_1}{r_0}, \qquad
    {\varrho_R} (V_{0} + V_1) = M^+  \, ({\mu_\infty - \mu_1}) \, \frac {r_1}{r^2_0}.
\end{aligned}
\end{equation}

Summarizing the results thus far: we are given the values of the remote chemical potential $\mu_\infty$, the inner radius $r_0$, the reference density $\varrho_R$ and the mobilities $M^\pm$. 
Up to now, we have three equations at our disposal, namely, \color{blue}(\ref{eq:v2-114MIT}), (\ref{eq:52})$_{1}$, and  (\ref{eq:52})$_{2}$, which involve five unknown quantities: 
\color{black}the chemical potentials $\mu_0, \mu_1$, the {accretion} velocities $V_0, V_1$, and the outer radius $r_1$. 
\color{blue}The extra two equations needed to close the system \color{black}will come from the kinetics of {accretion} at $\Sigma_0$ and $\Sigma_1$ which we turn to next.



\color{black}

\section{The kinetics of {accretion}.}   \label{sec-4}

In problems from continuum mechanics and materials science involving {accretion} of a body in the presence of deformation and mass transport, e.g.  the growth of a thin film in contact with a vapor reservoir of atoms, it is necessary to characterize the kinetics of the accretive prosess.
From a thermodynamic point of view {accretion} is in general a non-equilibrium process, and therefore involves a driving force (which is a measure of the departure from equilibrium) and a conjugate flux.  {Following Abeyaratne and Knowles \cite{RAJKK1990,RAJKK2000}} the appropriate driving force is determined by calculating the dissipation rate (or more generally the entropy production rate). Thermodynamic equilibrium corresponds to the vanishing of the driving force (often called the ``Maxwell condition''). One simple model of a kinetic law is a linear relation between the driving force and the conjugate flux, presumably appropriate for small departures from equilibrium. {For a general discussion of thermodynamics forces, conjugate fluxes and the kinetics of nonequilibrium processes the reader is referred to Chapter 14 of Kestin \cite{Kestin1979} or Chapter 14 of Callen \cite{Callen1985}.}

\medskip

\noindent {\bf Driving force and linear kinetics.}  
\noindent \color{blue}In the present setting, the total dissipation rate is
\[
\begin{split}
\Delta=&\phantom{{}+{}}\mbox{external mechanical power}
\\&+\mbox{inflow of chemical energy per unit time}
\\&-\frac{\rm d}{{\rm d}t}(\mbox{strain energy}).
\end{split}
\]
When body forces vanish, the only external mechanical power that must be accounted for is expended by the traction applied to the boundary of the body. Accordingly, we set
\begin{equation}\label{eq:61}
\mbox{external mechanical power}=\int_{\partial {\cal M}(t)} {\mathbf S}{\bf n}_R \cdot {\mathbf  V} \ dA.
\end{equation}
Here ${\bf n}_R$ is the outward unit normal on the boundary of the material manifold.  With regard to the velocity, it is important to note that the velocity of the boundary differs from the velocity of a particle that happens to be at the boundary.  The velocity of the boundary ${\bf V}$ is defined as follows:
consider a time-dependent  material point $\mathbf X_{\rm b}(t)$ which belongs to $\partial\mathcal M(t)$ and whose time derivative $\dot{\mathbf X}_{\rm b}$ is parallel to $\mathbf n_R$; then $\dot{\mathbf X}_{\rm b}=V\mathbf n_R$ with $V$ the outward velocity of $\partial\mathcal M(t)$; we let $\mathbf V=\frac{\rm d}{{\rm d}t}\bm\chi(\mathbf X_{\rm b}(t),t)$ or, equivalently, by the chain rule, 
\begin{equation}\label{eq:65}
\mathbf V=\mathbf v+V\mathbf F\mathbf n_R,
\end{equation}
where we recall that $\mathbf v=\dot{\bm\chi}$ is the velocity of the material point $\mathbf X_b(t)$. 

Next we write the external chemical power as the product of the chemical energy required to convert a unit mass of free particles into bound particles bounded to the body, multiplied by the rate at which mass is added to the body. In accordance with this notion, we set
\begin{equation}\label{eq:70}
\mbox{inflow of chemical energy per unit time}=\int_{\partial {\cal M}(t)} \varrho_R (\mu-\mu_R) V \ dA,
\end{equation}
where we interpret $\mu_R$ as the amount of energy needed to assemble a unit mass of solid material.

We now can write the \color{blue}total \color{black}dissipation rate as:
 \be
 \Delta = \int_{\partial {\cal M}(t)} {\mathbf S}{\bf n}_R \cdot \mathbf V \ dA   + \int_{\partial {\cal M}(t)}  \varrho_R(\mu-\mu_{R})V\color{black}\ dA 
 - \frac{d}{dt} \int_{{\cal M}(t)}  W(\mathbf F) \ dV. \label{20150609-eq52}
\ee
\color{black}By making use of standard divergence and transport theorems we rewrite the dissipation rate in the equivalent form
\be
\Delta = \int_{ \partial {\color{blue}\cal M}(t)} \left[\ {\bf S} {\bf n}_R \cdot   {\bf F}\mathbf n_R -   (W(\mathbf F) - (\mu-\mu_R) \varrho_R) \right]V\ dA .
\ee
Therefore in settings where surface {accretion} is the only nonequilibrium process we may identify
\be \label{eq-0620-dforce}
f = {\bf S} {\bf n}_R \cdot   {\bf F}{\bf n}_R -   (W(\mathbf F)- (\mu-\mu_R) \varrho_R) 
\ee
\color{blue}with
\color{black}the {\it driving force on the surface of {accretion}} and $V$ as its \emph{conjugate flux}. 
{An accretive process is characterized by a kinetic relation between the flux,  the driving force, and possibly other local fields: $V = \overline{V}(f, \ldots)$. In the simplest case, when the departure from thermodynamic equilibrium is small, one has a linear kinetic relation}\comment{Removed dependence on $\mathbf X$ and $t$. In fact $b$ may depend on $\mathbf X$ as well.}
\be \label{eq:0621-1}
\color{blue}f = b V
\ee
where the constant $b$ is a positive kinetic modulus. \color{blue}This is to hold at all points 
at which {accretion} occurs.\color{black} 

\color{blue2}The paper \cite{ZhuS1988continuum}, which focuses on a one-dimensional treadmilling structure, follows an approach similar to ours in the deduction of the evolution law governing boundary accretion. In particular, Eq. (5) in that paper is based on balance between dissipation, mechanical work, and supply of chemical energy. Alternatively, the equation governing accretion may be arrived at by making use of the notion of configurational or material force \cite{CiarlPM2013Mechanobiology,FriedG1999JSP,Gurti2000a}.\color{black}


 \noindent \textbf{Specialization to the problem at hand.}  At this point we make the following modeling choices pertaining to the boundary $\partial{\mathcal M}$ of the material manifold: 
\begin{itemize}
\item [(A7)] we {take the kinetic relation to be linear and} allow the kinetic moduli of the two parts $\Gamma_0$ and $\Gamma_1$ of the boundary  to be different; and
\item [(A8)] we allow the referential chemical potentials 
\color{blue}at \color{black} $\Gamma_0$ and $\Gamma_1$ to be different.
\end{itemize}
Accordingly we denote by $b_0 \, {(> \! 0)}, b_1 \, {(> \!0)}$ and $\mu_{R,0}, \mu_{R,1}$ the respective values of the kinetic modulus and referential chemical potential at $\Gamma_0, \Gamma_1$.    Thus we are distinguishing between the energetics and kinetics of the addition of material at $\Gamma_0$ and the removal of material at $\Gamma_1$. It is worth pointing out that, in the notation used here, the free particles have chemical potential $\mu_0$ just before they attach to the body and $\mu_{R,0}$ soon after; and likewise they have chemical potential $\mu_1$ soon after they detach from the body and $\mu_{R,1}$ just before. \textcolor{blue}{Thus, $\mu_{R,0} -\mu_{0}$ and $\mu_{1} - \mu_{R,1}$ are the respective changes in chemical energy during accretion and ablation.} \color{blue}We interpret the quantity $\varrho\mu_R$ as the energetic cost of adding a unit mass of material to the body, that is, the cost of accretion. As we shall see below, this extra energetic term substantially affects the evolution of growth. This point is also discussed in the paper \cite{tiero2014morphoelastic}, which contains other examples on how the cost of accretion may be relevant to the kinetics of growth.\color{black}

In the specific problem at hand, we recall that the outward unit normal $\mathbf n_R$ to $\partial{\mathcal M}$ is (\emph{cf.} Figure 2):
\begin{equation}\label{eq:32ra}
\mathbf n_R(\mathbf X,t)=
\begin{cases}
-\mathbf e_Z\qquad\textrm{for} \ \mathbf X\in\Gamma_0(t),\\
\phantom{-}\mathbf e_Z\qquad\textrm{for} \ \mathbf X\in\Gamma_1(t).
\end{cases}
\end{equation}
As encountered previously in  \eqref{eq:39} and (\ref{eq:1MITv2}), the outward normal velocities of points of the boundaries $\Gamma_0$ and $\Gamma_1$ are
$V_0 {\mathbf  n}_R$ and $V_1 {\mathbf  n}_R$. Therefore
we take the kinetic equations at $\Gamma_0$ and $\Gamma_1$ to be
 \be \label{eq:0621-11}
 f_0 = b_0 V_0, \qquad f_1 = b_1 V_1,
 \ee
 where, as we show below, the driving forces, $f_0$ and $f_1$, on the respective boundaries $\Gamma_0$ and $\Gamma_1$ of the material manifold, are
\begin{subequations}
\begin{align}
   &\displaystyle f_0 =  (\mu_{0}- \mu_{R,0}) \varrho_R - W\big( {r_1}/{r_0} \big), \qquad {\rm and} \label{eq:38ra}\\[2ex]
   &\displaystyle f_1 = (\mu_{1}-\mu_{R,1})\varrho_R - W\big({r_1}/{r_0}\big). \label{eq:36ra}
\end{align}
\end{subequations}
Here $W(\lambda)$
is the restriction of the strain energy function $\widehat{W}(\lambda_1, \lambda_2, \lambda_3)$ to  isochoric equi-biaxial deformations as introduced previously in  (\ref{20150114.rhotel11}).

It is noteworthy  that, even though the same term $W( {r_1}/{r_0} )$ appears in both (\ref{eq:38ra}) and  (\ref{eq:36ra}), 
\color{blue}it originates
\color{black}from different 
\color{blue}contributions
\color{black}in the general expression for driving force: in the expression for the driving force $f_0$ on $\Gamma_0$, it appears from the first term in (\ref{eq-0620-dforce}), the term related to stress; see \eqref{eq:21}. On the other hand in the expression for the driving force $f_1$ on $\Gamma_1$, it appears from the second term in (\ref{eq-0620-dforce}) related to the free energy.

\begin{proof}[Derivation of the expression \eqref{eq:38ra} for the driving force on $\Gamma_0$:]
By \eqref{eq:28} and \eqref{eq:4}, the principal stretches at $\Gamma_0$ are $\lambda_r(r_0)=\lambda_\theta(r_0)=1$ and hence 
\begin{equation}\label{eq:45ra}
\mathbf F \big|_{\Gamma_0}=\mathbf{e}_\theta\otimes \mathbf{e}^\Theta+\mathbf{e}_\phi\otimes \mathbf{e}^\Phi+\mathbf{e}_{r}\otimes \mathbf{e}^Z.
\end{equation}
The strain energy at $\Gamma_0$, $\widehat{W}(1,1,1)$, vanishes, and the free energy therefore contains only a chemical contribution:
\begin{equation}\label{eq:48ra}
  \psi|_{\Gamma_0}=\mu_{R,0}\varrho_R.
\end{equation}
Next, from  \eqref{eq:45ra} and \eqref{eq:32ra}, 
  \begin{equation}
    (\mathbf F\mathbf n_R)\big|_{\Gamma_0}=\mathbf F\big|_{\Gamma_0}(-\mathbf e_Z)=\mathbf e_r,
  \end{equation}
and by making use of \eqref{eq:21},  ${\bf S} = \det {\bf F} \, {\bm\sigma} {\bf F}^{-T}$ and $\det \, {\bf F} = 1$, we find that
\begin{equation}\label{eq:49ra}
  (\mathbf F\mathbf n_R\cdot {\bf S}\mathbf n_R)\big|_{\Gamma_0}=(\mathbf e_r\cdot \bm \sigma\mathbf e_r)\big|_{\Sigma_0} =\sigma_r(r_0,t)= - W(r_1/r_0).
\end{equation}
Then substituting \eqref{eq:48ra} and \eqref{eq:49ra} into \eqref{eq-0620-dforce} gives \eqref{eq:38ra}.
\end{proof}

\begin{proof}[Derivation of the expression \eqref{eq:36ra} for the driving force on $\Gamma_1$:]
It is readily seen from $(\ref{eq:28})_2$ and \eqref{eq:33} that
\begin{equation}
  \lambda_r(r_1)=  {r^2_0}/{r^2_1},\qquad \lambda_\theta(r_1)={r_1}/{r_0},
\end{equation}
and so from (\ref{20150114.rhotel11})
\begin{equation}\label{eq:50ra}
  \widehat W(\lambda_1, \lambda_2, \lambda_3)\big|_{\Gamma_1} = W(r_1/r_0).  
\end{equation}
Since the surface $\Gamma_1$ is traction free, we have
\begin{equation}\label{eq:51ra}
   (\mathbf F\mathbf n_R\cdot {\bf S}\mathbf F\mathbf n_R)\big|_{\Gamma_1}=(\mathbf e_r\cdot \bm\sigma\mathbf e_r)\big|_{\Sigma_1}=\sigma_r(r_1,t)=0,
\end{equation}
again having used ${\bf S} = \det {\bf F} \, {\bm\sigma} {\bf F}^{-T}$ and $\det \, {\bf F} = 1$.
Substituting \eqref{eq:50ra} and \eqref{eq:51ra} into \eqref{eq-0620-dforce} yields \eqref{eq:36ra}.
\end{proof}

 \medskip
 

Finally, by substituting the expressions (\ref{eq:38ra}) and (\ref{eq:36ra}) for the driving forces into the kinetic equations (\ref{eq:0621-11}) we arrive at the pair of equations
\begin{subequations}
\begin{align}
   &\displaystyle b_{0}V_{\rm 0}=(\mu_{0}-
\mu_{R,0}) \varrho_R-W\left({r_1}/{r_0}\right),\label{eq:38}\\[1ex]
   &\displaystyle b_{1}V_{1}=(\mu_{1}-\mu_{R,1})\varrho_R-W\left({r_1}/{r_0}\right),\label{eq:36}
\end{align}
\end{subequations}

\color{blue}  

\noindent \textbf{Remark:} Recall that the strain energy function $W(\lambda)$ defined in (\ref{20150114.rhotel11}) and appearing above is increasing for $\lambda\ge 1$; see (\ref{20150617.v2777}).
Recall also from the discussion below  \eqref{eq:38ra},  \eqref{eq:36ra} that the $W$ term in \eqref{eq:38} enters via the stress and so we may conclude that stress always hinders {accretion} at the inner surface.  On the other hand we observed in that same discussion that the $W$ term in \eqref{eq:36} enters via the strain energy, not stress, and this shows that strain energy promotes {ablation} at the outer surface. Since $W \geq 0$ and $V_0 >0$ we see from \eqref{eq:38} that necessarily
\begin{equation}
\mu_0 > \mu_{R,0}.
\end{equation}
There is no similar requirement at the outer surface.


In the summary at the end of Section \ref{sec-3MIT} we observed that two more equations were needed in order to solve the problem stated there.  These are provided by (\ref{eq:38}) and (\ref{eq:36}) and so we are now in a position to carefully state the problem of interest and to analyze it. We turn to this next. \color{black}


\section{ \textcolor{blue}{The treadmilling regime: analysis and results.}}\label{sec-5}

\subsection{The system governing treadmilling.}

In the treadmilling regime all evolution processes are steady and so it follows from \eqref{eq-20151128-2} and \eqref{eq:52} that the velocities $V_0$ and $V_1$ are necessarily constant.  Therefore from \eqref{eq:0621-11},  the driving forces $f_0$ and $f_1$ must also be constant. Not surprisingly, it now follows from \eqref{eq:38ra} (or \eqref{eq:36ra}) and \eqref{20150617.v2777} that
 the outer radius $r_1$ of the body, which in general is time dependent due to {accretion}, remains constant in the treadmilling regime:
\begin{equation}
\dot{r}_1 = 0.
\end{equation}
It is immediate from (\ref{eq:v2-114MIT}) that 
\begin{equation} \label{eq:53}
  V_{1}= -V_0.
\end{equation}
This too is not surprising \color{black}since under stationary conditions the addition of material at $\Gamma_0$ will be balanced by its removal at $\Gamma_1$.
Observe now from \eqref{eq:52}$_2$ that, because of (\ref{eq:53}),
\be
\mu_1 = \mu_\infty.
\ee
Equation (\ref{eq-20151128-2})$_2$ tells us that in the treadmilling regime $\mu(r)= \mu_\infty$ for $ r \geq r_1$, and therefore that there is no {free particle flux} outside of the solid body.  This reflects the fact that in the treadmilling regime the accretive process is self-sustaining in the sense that the {mass of free particles} being attached to the body at $\Sigma_0$ is precisely equal to the {mass of free particles} detaching from it at $\Sigma_1$.
Since the mobility $M^+$ outside the solid no longer affects the analysis, it is convenient to set
\begin{equation}
M = M^-
\end{equation}
from hereon.

We can now eliminate $\mu_1$ and $V_1$ and reduce the problem to solving the \emph{treadmilling system} consisting of the three equations
\begin{subequations}\label{eq:55}
\begin{align}
 {\varrho_R} V_{0}&= M \, \frac{\mu_\infty-\mu_0}{r_1-r_0}\frac {r_1}{r_0},\label{eq:56}\\
\displaystyle b_{0}V_{\rm 0}&=(\mu_{0}-
\mu_{R,0}) \varrho_R - W\left({r_1}/{r_0}\right),\label{eq:57}\\[1ex]
\displaystyle b_{1}V_{0}&=-(\mu_{\infty}-\mu_{R,1})\varrho_R + W\left({r_1}/{r_0}\right),\label{eq:58}
\end{align}
\end{subequations}
for the remaining unknowns $(V_0,\mu_0,r_1)$ with 
\begin{equation}\label{eq:23}
\text{$V_0 > 0$ and $r_1 > r_0$.}
\end{equation}
Recall that \eqref{eq:56} follows from Fick's law combined with balance of mass relating the {free particle flux} reaching $\Sigma_0$ per unit time with the rate of {accretion} of ${\mathcal M}$ at $\Gamma_0$. 
The remaining equations follow from the kinetic laws at the {accretion} fronts $\Gamma_0$, $\Gamma_1$.\medskip


\subsection{\color{blue}Results.} \label{subsec-results}
\color{blue} In this subsection we state the
main analytical results of this study: \color{black} $(i)$ the solvability of the 
system governing treadmilling\color{black}; and $(ii)$ 
calculate asymptotic estimates for the thickness of the {solid} and the rate of {accretion} when the bead radius is either much smaller than or much larger than a characteristic length. \color{blue} Proofs of these results are given in the next subsection, while several implications are discussed in Section \ref{sec-8}.\color{black} 



%

  It will be useful in what follows to let
 $V_*, V_{**}$ and $\ell_*$  denote the following characteristic velocity and length scales:
\begin{equation} \label{eq:RA20151113-1}
 V_*:=\frac{\mu_{R,1}-\mu_{R,0}}{b_0+b_1}\varrho_R,  \qquad V_{**}:=\frac{\mu_{R,1}-\mu_{\infty}}{b_1}{\varrho_R}, \qquad \ell_*:= \frac{(b_0+b_1)M}{\varrho^2_R}.
 \end{equation}
The terms on the right hand sides of these equations are all known and so we may consider $V_*, V_{**}$ and $\ell_{*}$ to be given. We will see shortly that the {accretion} velocity lies in the range $V_{**} < V_0 < V_*$ and asymptotic results will derived when  $r_0/\ell_* \to 0$ or $\infty$.

\begin{proposition}[Solvability of the treadmilling system]
The treadmilling system \eqref{eq:55} admits a solution $(V_0,\mu_0,r_1)$ satisfying $V_0>0$ and $r_1>r_0$ if and only 
\begin{equation} \label{eq:0622-2}
V_* > 0
\end{equation}
and
\begin{equation}\label{eq:29}
V_* > V_{**}.
\end{equation}
Moreover, when \eqref{eq:0622-2} and  \eqref{eq:29} hold, this solution is unique.  Furthermore, the  {accretion} velocity $V_0$ in the treadmilling regime satisfies
\begin{equation}\label{eq:24}
  V_{**}<V_0<V_*.
\end{equation}
\end{proposition}

The functions $V_0 = V_0(\eta), \mu_0 = \mu_0(\eta)$ and $r_1 = r_1(\eta)$ that solve the treadmilling system depend continuously on the nondimensional bead radius
\begin{equation}\label{eq:27}
  \eta:=\frac{r_0}{\ell_*}.
\end{equation}
Moreover it can be shown that these functions have finite limiting values when $\eta \to 0$ and $\infty$.  The next three propositions are concerned with the asymptotic behavior of the functions $V_0(\eta)$ and $d(\eta) = r_1(\eta) - r_0$ in those limits,
 the latter being the  \emph{thickness} 
\begin{equation}\label{eq:48}
  d:=r_1-r_0
\end{equation}
of the solid.  The conditions for treadmilling established in Proposition 1 are assumed to hold and so the existence of a unique solution is taken for granted.

\begin{proposition}\label{prop:2} {\rm (Small bead.)} 
When the nondimensional bead radius $\eta \to 0$ the thickness of the solid $d(\eta)$  and the {accretion} velocity $V_0(\eta)$ have the limiting values
\begin{equation}\label{eq:59}
  \frac{d(\eta)}{r_0}\to \nu_*-1, \qquad V_0(\eta) \to V_*,
\end{equation}
where $\nu_*>1$ is the unique root of
\begin{equation}\label{eq:31}
 \frac{1}{b_1} W(\nu_*) = V_* - V_{**}.
\end{equation}
\end{proposition}

\begin{proposition}\label{prop:3} {\rm (Large bead. Case $V_{**}\geq 0$.)}
When the nondimensional bead radius $\eta \to \infty$ with $V_{**} \geq 0$, the thickness of the solid has the limiting value
\begin{equation}\label{eq:54}
  \frac {d(\eta)} {r_0}  \to 0,
\end{equation}
with the asymptotic form
\begin{equation}\label{eq:541121}
  \frac {d(\eta)} {r_0} \sim  \left( \frac {V_*}{V_{**}}-1\right) \ \frac 1 \eta. 
\end{equation}
The {accretion} velocity 
\begin{equation}\label{eq:541122}
V_0(\eta) \to V_{**} 
\end{equation}
in this limit.
\end{proposition}

\begin{proposition}\label{prop:4} {\rm (Large bead.  Case $V_{**}<0$.)}
When the nondimensional bead radius $\eta \to \infty$ with $V_{**} <0$, the thickness of the solid 
\begin{equation} \label{eq:68}
\frac {d(\eta)}{r_0}\to \nu_{**}-1 , 
\end{equation}
where
$\nu_{**} \textcolor{blue}{>1}$ is the unique root of
\begin{equation}\label{eq:66}
\frac{1}{b_1}   W(\nu_{**})= - V_{**}.
\end{equation}
The {accretion} velocity has the limiting value
\begin{equation} \label{eq:681121}
V_0(\eta) \to 0,
\end{equation}
with the asymptotic form
\begin{equation}
V_0(\eta) \sim  \frac{V_*}{1-1/\nu_{**}} \ \frac 1 \eta
\end{equation}
as $\eta \to \infty$.

\end{proposition}


\subsection{Proofs of results.} \label{subsec-proofs}

{It is convenient to rewrite the treadmilling system \eqref{eq:55} in the following equivalent form}
 \begin{subequations}\label{20151119-123}
\begin{align}
 \displaystyle &\frac{V_*}{1 + \eta (1 - 1/\nu)} = V_{**} + \frac{1}{b_1} W(\nu), \label{20151119-5}\\
 &V_0 = V_{**} + \frac{1}{b_1} W(\nu),\label{20151119-1}\\
\displaystyle &(b_0 + b_1)(V_* - V_0) = (\mu_\infty - \mu_0)\rho_R,\label{20151119-2}
\end{align}
\end{subequations}
where we have used \eqref{eq:RA20151113-1} and let
$$
\nu=r_1/r_0.
$$
Equation  \eqref{20151119-5} is obtained by eliminating $\mu_0$ and $V_0$ from  \eqref{eq:55};  \eqref{20151119-1} is equivalent to \eqref{eq:58};  and \eqref{20151119-2} is obtained by adding  \eqref{eq:57} and  \eqref{eq:58}.

The only unknown in equation \eqref{20151119-5}  is $\nu$. If it can be solved for $\nu(\eta)$  then \eqref{20151119-1} gives $V_0(\eta)$ and \eqref{20151119-2} gives $\mu_0(\eta)$.

\noindent {\bf Proposition 1:} (Solvability of the treadmilling system.) Define the functions\footnote{\color{blue}We use the subscript $\eta$ on $g_\eta$ to explicitly display its dependence on $\eta$ since we shall be examining the limiting cases $\eta \to 0$ and $\infty$ in the subsequent propositions.}:
\begin{equation} \label{eq:16}
\begin{aligned}
  &g_\eta(\lambda):= \frac{V_*}{1+\displaystyle\eta \left(1-  1/ \lambda\right)},\\
 &h(\lambda):=V_{**} + \frac{1}{b_{1}}{W\left(\lambda\right)},
\end{aligned}
\end{equation}
for $\lambda \geq 1$. 
By making use of \eqref{20151119-1} and \eqref{20151119-5} we see that the solvability of the treadmilling problem is equivalent to there being roots $V_0 > 0, \, \nu > 1$ of the system
\begin{align} \label{eq:0622-1}
V_0=g_\eta(\nu),\qquad V_0=h(\nu).
\end{align}

We are now in a position to establish Proposition 1. First, observe from (\ref{eq:16})$_1$ that $g_\eta(\lambda)$ has the same sign as $V_*$ for all $\lambda \geq 1$. Since it is required that $V_0 >0$, it now follows because of \eqref{eq:16}$_1$ and (\ref{eq:0622-1})$_1$  that it is necessary that $V_* >0$.

Second, granted $V_* >0$, it can be readily confirmed that the continuous function $g_\eta(\lambda)$ is monotonically decreasing for $\lambda \ge 1$ and satisfies:
$$
g_\eta(\lambda)\le g_\eta(1)=V_* \ \text{for all $\lambda\ge 1$} \qquad \text{and} \qquad \lim_{\lambda \to+\infty}
g_\eta(\lambda)= V_*/(1 + \eta).
$$
Likewise, it is readily seen from the definition of $W(\lambda)$ given in \eqref{20150114.rhotel11}, together with (\ref{20150617.v2777}), that the continuous function $h(\lambda)$ is monotonically increasing for $\lambda > 1$ and satisfies:
$$
 h(\lambda) \ge h(1)=V_{**}  \  \text{for all $\lambda\ge 1$} \qquad \text{and} \qquad \lim_{\lambda \to+\infty}h(\lambda)=+\infty.
$$
It now follows that the equation $g_\eta(\nu) = h(\nu)$ has no root $\nu >1$ unless $V_{**} < V_*$ which is a second necessary condition for there to be a solution.

Conversely when $V_* >0$ and $V_{**} < V_*$, the foregoing considerations show that the system  \eqref{eq:0622-1} has a unique solution $V_0 >0$ and that it lies in the range $V_{**} < V_0 < V_*$.  Proposition 1 is thus established. Figure 3 shows schematically the graphical construction to find the solution of the treadmilling system. The graphs of $g_\eta(\lambda)$ and $h(\lambda)$ intersect at only one point, whose coordinates deliver the solution to system \eqref{eq:0622-1}. 

\begin{figure}[h]
\centerline{\includegraphics[scale=.85]{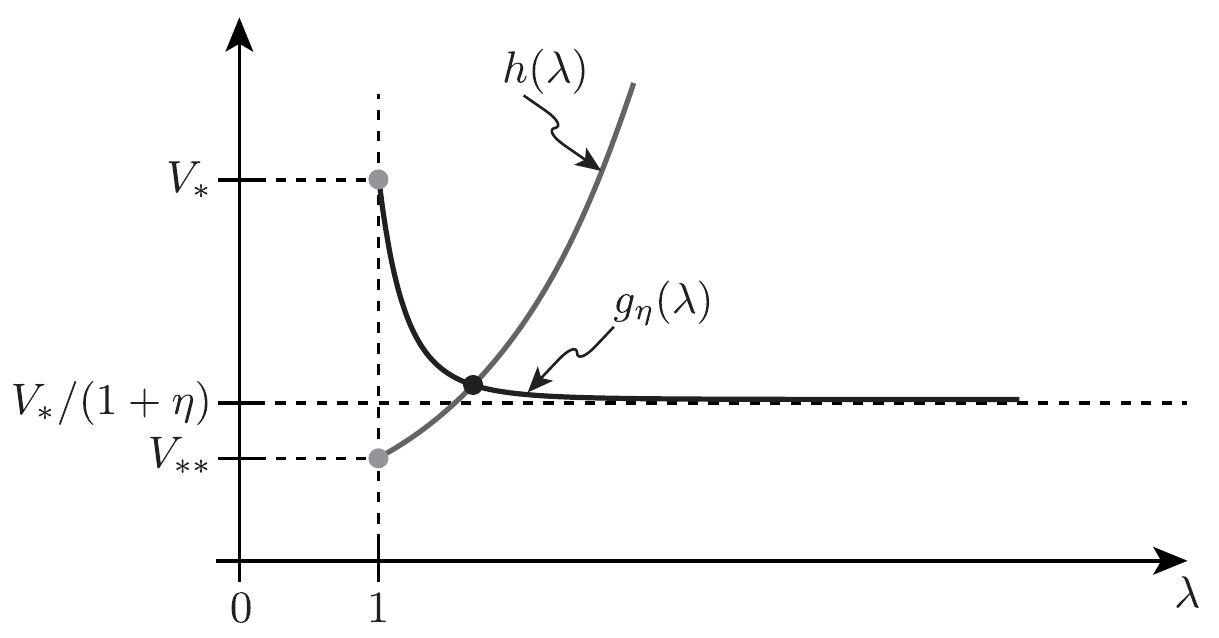}}

\smallskip
\caption{\footnotesize Schematic plots of the functions $g_\eta(\lambda)$ and $h(\lambda)$ defined in \eqref{eq:16}: $g_\eta$ decreases monotonically from the value $g_\eta(1)=V_*$ while $h$ increases monotonically from the value $h(1)=V_{**}$. As $\lambda \to \infty$, $h$ becomes unbounded while $g$ converges to $V_*/(1+\eta) >0$. The coordinates of the unique point of intersection are $(r_1/r_0, V_0)$. The figure has been drawn for the case $V_{**} >0$ though it is possible for $V_{**}$ to be negative.} \label{Fig-1_DB.pdf}
\end{figure}

\begin{figure}[h]
\centerline{\includegraphics[scale=.85]{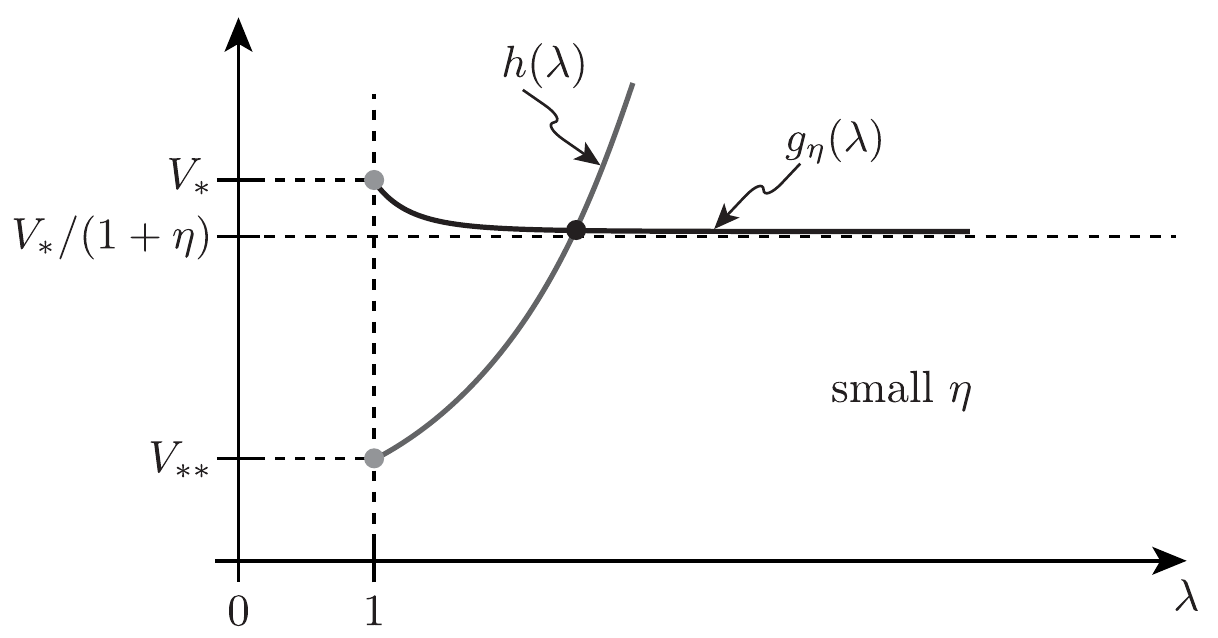}}

\smallskip
\caption{\footnotesize Schematic plots of the functions $g_\eta(\lambda)$ and $h(\lambda)$ for a small value of $\eta$. As $\eta \to 0$ we have $g_\eta \to V_*$ at each $\lambda$.  Since the coordinates of the unique point of intersection are $(r_1(\eta)/r_0, V_0(\eta))$,  we expect that $V_0(\eta) \to V_*$ in the limit $\eta \to 0$.} \label{Fig-2_DB.pdf}
\end{figure}


\begin{figure}[h]
\centerline{
\raisebox{6ex}{\includegraphics[scale=0.7]{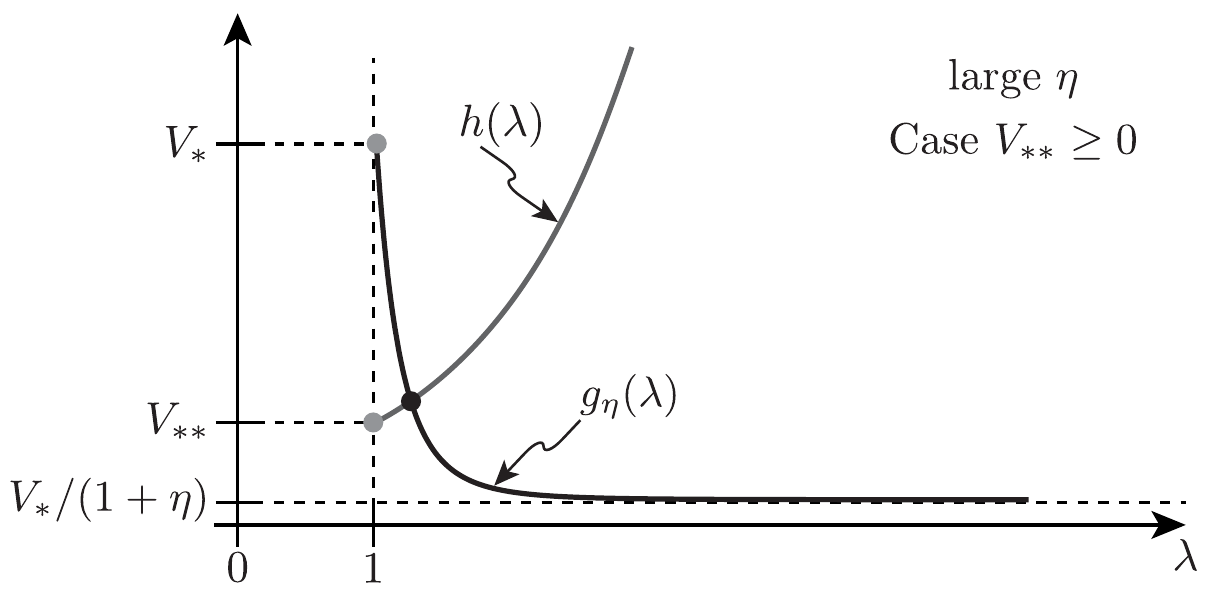}}\quad 
\includegraphics[scale=0.7]{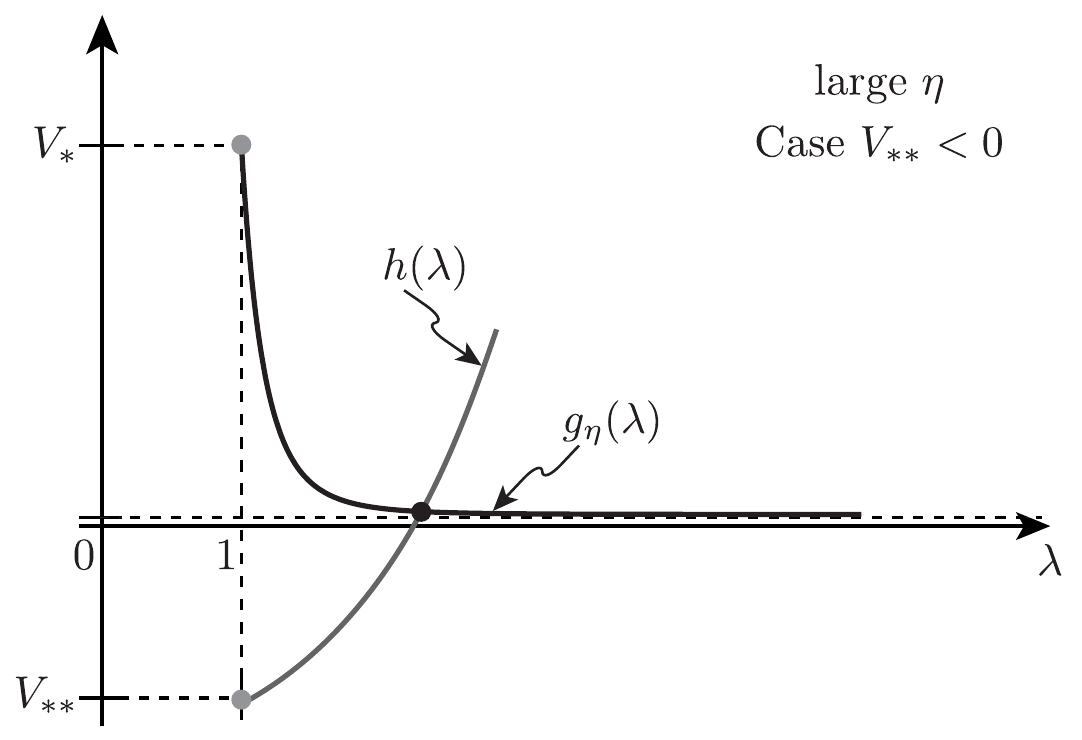}
}
\smallskip
\caption{\footnotesize Schematic plots of the functions $g_\eta(\lambda)$ and $h(\lambda)$ for a large value of $\eta$. The graph of $g_\eta$ starts from the point $(1, V_*)$ and declines rapidly towards the value $V_*/(1+\eta)$.  As $\eta \to \infty$ the function $g_\eta \to 0$ at each $\lambda>0$ while $g_\eta(1) = V_*$. The coordinates of the unique point of intersection are $(r_1(\eta)/r_0, V_0)$.  Therefore in the case $V_{**} \geq 0$ (left) we expect that $r_1(\eta)/r_0 \to 0$ as $\eta \to \infty$.  In the case $V_{**} <0$ (right), we expect $V_0(\eta) \to 0$ as $\eta \to \infty$. } 
\label{Fig-1_DBRA2.pdf}
\end{figure}


We now turn to the propositions concerning the behavior of the solution $V_0(\eta), r_1(\eta), \mu_0(\eta)$ in the limits $\eta \to 0$ and $\eta \to \infty$. We can anticipate the results by the following graphical discussion. Observe from  \eqref{eq:16} that the function $h(\lambda)$ does not depend on $\eta$ but $g_\eta(\lambda)$ does.  First, when $\eta \to 0$, we see that $g_\eta \to V_*$ at each fixed $\lambda$.  Since the point at which the two curves intersect has coordinates $(r_1(\eta)/r_0, V_0(\eta))$, the schematic Figure \ref{Fig-2_DB.pdf} informs us that $V_0(\eta) \to V_*$ in this limit. In the other limit when $\eta \to \infty$ we see from \eqref{eq:16} that $g_\eta(\lambda) \to 0$  for $\lambda >1$ with $g_\eta(1) = V_*$. Thus for a large value of $\eta$, the function $g_\eta$ decreases rapidly from the value $V_*$  towards the value $V_*/(1+\eta)$ as $\lambda$ increases,  and so the graph of $g_\eta$ has a boundary layer near $\lambda = 1$ as indicated in the schematic plots depicted in Figure \ref{Fig-1_DBRA2.pdf}. Again, since the point at which the two curves intersect has coordinates $(r_0(\eta/r_1, V_0(\eta))$, the figures indicate that $V_0(\eta) \to 0$ if $V_{**} < 0$ and $r_0(\eta)/r_1 \to 1$ if $V_{**} > 0$.

We now turn to the analysis of these limiting cases. The necessary and sufficient conditions, $V_* > 0, \, V_* > V_{**}$, for the existence of a unique solution to the treadmilling problem are assumed to hold. To this effect, we observe that an application of Dini's implicit function theorem to equation \eqref{20151119-5}, which defines $\nu(\eta)$ implicitly, \color{blue} ensures \color{black}  that $\nu(\eta)$ is continuous and strictly decreasing for $\eta\ge 0$.\color{black}

\noindent {\bf Proposition 2:}  (Small bead.) \ 
Since $\nu(\eta)$ is continuous, we have: 
\begin{equation}\label{eq:35}
\nu(\eta) \to \nu_* \geq 1 \qquad {\text{as}} \quad \eta \to 0. 
\end{equation}
The limit $\nu_*$ can be identified by taking the limits of both sides of the equation \eqref{20151119-5}, which yields
\be \label{20151119-6}
V_*= V_{**} + \frac{1}{b_1} W(\nu_*),
\ee
an equation that, \color{blue} by a special case of Propoisition 1, \color{black}  is guaranteed to have a unique solution $\nu_* >1$. Equation  \eqref{20151119-1} together with \eqref{20151119-6} now gives 
\be
V_0(\eta) \to V_*,
\ee
and equation \eqref{20151119-2} gives
\be
\mu_0(\eta) \to \mu_\infty.
\ee
This establishes Proposition 2.

We now turn to the limit $\eta \to \infty$.  Suppose that 
$$
\nu(\eta) \to \nu_{**} \geq 1 \qquad {\rm as} \quad \eta \to \infty. 
$$ 
We see from the denominator of the left hand side of \eqref{20151119-5} that we have to distinguish between the cases $\nu_{**} =1$ and $\nu_{**} >1$.

\noindent{\bf Proposition 3:} (Large bead. Case $V_{**} \ge 0$.) 
Since $\nu(\eta)$ is decreasing, and since $\nu(\eta)\ge 1$, we have that $\nu(\eta)$ converges to some $\nu_{**}\ge 1$ as $\eta\to\infty$. The possibility that $\nu_{**}$ is strictly greater than $1$ \color{blue}  can be ruled out since, otherwise, on passing to the limit in \eqref{20151119-5} we would obtain $0=V_{**} + W(\nu_{**})/b_1$, which cannot hold since $V_{**}\ge 0$ and $W(\lambda)>0$ for $\lambda>0$. Thus, we conclude that:\color{black}
\be
\nu(\eta) \to \nu_{**}=1\qquad {\rm as} \quad \eta \to \infty. 
\ee
Then equation  \eqref{20151119-1} together with $W(\nu(\eta)) \to W(1)=0$ gives 
\be\label{eq-20151204-1}
V_0(\eta) \to V_{**}.
\ee
Since we need $V_0 >0$ it follows that this case occurs only if $V_{**} \geq 0$ which is precisely the condition under which Proposition 3 has been claimed to hold.
Using the limiting value of $V_0$ from \eqref{eq-20151204-1}  in equation \eqref{20151119-2} gives
\be
\mu_0(\eta) \to  \mu_\infty  + \frac{b_0 + b_1}{\rho_R} (V_{**} - V_*).
\ee
Note  from \eqref{20151119-5} and $W(\nu(\eta)) \to 0$ that 
$$
\big( \nu(\eta) - 1 \big) \eta \to  \frac{V_* - V_{**}}{V_**}  \qquad {\rm as } \ \eta \to \infty,
$$
which we can write in the form of the asymptiotic estimate 
$$
\nu(\eta)  \sim 1 + \frac{V_* - V_{**}}{V_**} \ \frac 1 \eta.
$$
This establishes  Proposition 3.

\noindent{\bf Proposition 4:} (Large bead. Case $V_{**} < 0$.)  We now argue that
\be
\nu(\eta) \to \nu_{**} >1 \qquad {\rm as } \ \eta \to \infty.
\ee
\color{blue} As in Proposition 3, $\nu(\eta)$ converges monotonically to a limit $\nu_{**}\ge 1$. However since $g_\eta$ is a positive function, we must also have $h(\nu_{**})>0$, and this is only possible if $\nu_{**}>1$, since we are in the case $V_{**}<0$ (see Figure \ref{Fig-1_DBRA2.pdf}).\color{black} 

In this case, when $\eta \to \infty$ equation \eqref{20151119-5} gives
\be \label{eq-20151213-1}
V_{**} + \frac{1}{b_1} W(\nu_{**}) = 0.
\ee
Again, as noted in the preceding subsection, in view of the properties of $W$, \eqref{eq-20151213-1} has a unique root $\nu_{**} >1$ provided that $V_{**} < 0$.  This is precisely the condition under which Proposition 4 was stated to hold.  Equation \eqref{20151119-1} now tells us that
\be
V_0(\eta) \to 0
\ee
and \eqref{20151119-2} tells us that
\be
\mu_0(\eta) \to \mu_\infty - \frac{b_0 + b_1}{\rho_R} V_* = \mu_\infty + \mu_{R,0} - \mu_{R,1}.
\ee
Observe that \eqref{20151119-5} and \eqref{20151119-1} that
\be
\eta V_0(\eta) \to \frac{V_*}{1 - 1/\nu_{**}}  \qquad {\rm as } \ \eta \to \infty
\ee
which we can write in the form of the asymptotic estimate
\be
V_0(\eta) \sim \frac{V_*}{1 - 1/\nu_{**}} \ \frac 1 \eta.
\ee
This establishes Proposition 4.

\color{black}

\section{Discussion and \textcolor{red}{summary}} \label{sec-8}

We now examine various implications of the results of the previous section, and make some remarks on their place within a larger perspective.

 \medskip


 According to Proposition 1, necessary and sufficient conditions for the existence of a unique solution to the treadmilling porblem are that $V_* > 0$ and $V_* > V_{**}$.  These conditions can be written more illuminatingly in terms of the chemical potentials by using \eqref{eq:RA20151113-1} in the respective forms
 \be \label{eq:0622-2a}
\mu_{R,1} > \mu_{R,0},
\ee 
and
\begin{equation} \label{eq:RA1}
\mu_\infty > \mu_* \ := \frac{b_0 \mu_{R,1} + b_1 \mu_{R,0}}{b_0 + b_1}.
\end{equation}
The former inequality states that the referential chemical potential at the outer surface must exceed that at the inner surface, while the latter requires the remote chemical potential $\mu_\infty$ to
exceed a certain ``mean chemical potential'' $\mu_*$.

{Accretion} at the inner surface is limited by both diffusion and stress build-up. The former is characterized by the terms in \eqref{eq:55} that involve differences in chemical potential; the latter by the terms involving $W$.  It is seen from \eqref{eq:56} that $\mu_\infty - \mu_0$ increases as $d$ increases (at constant $r_0, M$ and $V_0$).  This implies, as one would expect, that  the larger  the thickness $d$, the larger is the chemical potential drop $\mu_\infty-\mu_0$ necessary to support a given {accretion} rate $V_0$.
 We also see from \eqref{eq:57} that the smaller the value of the chemical potential $\mu_0$ at the inner surface, the slower the {accretion} rate $V_0$ will be (at the same values of the other terms in \eqref{eq:57}).

The effects of mechanics on the {accretion} rate  are lumped into the last term of the right-hand side of \eqref{eq:57}.  When the thickness $d$ increases, so does the ratio $r_1/r_0$, and therefore, since $W(\lambda)$ is an increasing function for $\lambda > 1$, it follows from \eqref{eq:57} that   the {accretion} velocity $V_0$ decreases. 
As for the {ablation} rate at the outer surface, as noted previously, the discussion below  \eqref{eq:38ra},  \eqref{eq:36ra}, shows that the $W$ term in \eqref{eq:36} enters via the strain energy, not stress, and this shows that strain energy promotes {ablation} at the outer surface. 


\textbf{The limiting cases of small and large beads:}   The respective limiting values of the ratio $r_1/r_0$ when $\eta \to 0$ and $\eta \to \infty$ (with $V_{**} < 0$), i.e. $\nu_*$ and $\nu_{**}$, are given by the roots of  \eqref{eq:31} and  \eqref{eq:66}.  In view of \eqref{eq:29}, the right-hand side of \eqref{eq:31} is {positive}. Thus, given the properties \eqref{20150617.v2777}, \eqref{eq:RA2} of $W(\lambda)$, there is {precisely one} root $\nu_*>1$ of \eqref{eq:31}.  Similar considerations apply to \eqref{eq:66}. 
\medskip


 \textbf{Case of a small bead:}  According to Proposition \ref{prop:2}, we may take
  \begin{equation}\label{eq:63}
    \frac {d} {r_0}\simeq \nu_*-1, 
  \end{equation}
as an approximate formula for the thickness of the \textcolor{blue}{solid} when $r_0<<\ell_*$.  Here $\nu_* > 1$ is the unique root \eqref{eq:31}.  
Although \eqref{eq:31} cannot, in general, be solved explicitly for $\nu_*$, we can do so if we replace the function $W(\lambda)$ by its second-order Taylor expansion at $\lambda=1$.  On using the result in \eqref{eq:63} we obtain the following estimate for the thickness in this regime:
\begin{equation}\label{eq:62}
  \frac {d} {r_0} \simeq  \sqrt{\frac{2 (\mu_\infty  - \mu_{*}) \varrho_R}{W''(1)}},  
\end{equation}
which is expected to be accurate when $\mu_\infty - \mu_*$ is small.\comment{\color{red}GT: Can we spend a few extra words on explaining why the accurateness is contingent on having $\mu_\infty - \mu_*$ small? RA: This is because we replaced $W(\lambda)$ by its quadratic approximation which would be good for $\lambda \approx 1$. This means $\nu_* \approx 1$ and so $h$ small. Not sure how to say this without disrupting the flow.}

 Equation \eqref{eq:62} shows in particular that the thickness $d$ is proportional to the radius $r_0$. This agrees with the results of  \cite{Noir2000}, whose formula (26) gives $d \simeq r_0\sqrt{\Delta\widetilde\mu/Ca\xi^2}$, where $\Delta\widetilde\mu$ is the chemical energy released in the {polymer}ization process, $C$ is the elasticity modulus of the actin gel, $a$ is the distance between two \textcolor{blue}{actin units} in an F-actin chain, and $\xi$ is the average distance between nucleating proteins on the surface of the bead; this estimate was obtained by imposing a balance between chemical energy gained and the elastic energy expended during \textcolor{blue}{polymerization}.  A comparison with our results is straightforward if we identify $\mu_\infty - \mu_*$ with $\Delta\widetilde\mu$ and  observe that $\varrho_R \propto 1/(a\xi^2)$ and $C\simeq W''(1)$.

As can be seen from \eqref{eq:56}, \eqref{eq:48} and \eqref{eq:59},  $\mu_0(\eta) \to \mu_\infty$ when $\eta \to 0$.  Because of this, and since $\ell_* \propto M$ by \eqref{eq:RA20151113-1},  the limit $\eta \to 0$ is associated with the diffusion constant  $M \to \infty$ and the chemical potential drop  $\mu_\infty - \mu_0 \to 0$.  In this regime it is the stress build-up that inhibits \textcolor{blue}{accretion} at the inner surface.  We therefore refer to this as the \emph{stress-limited regime}. Since  $\nu(\eta) \to \nu_* > 1$ when $\eta \to 0$, the thickness $d(\eta) = (\nu(\eta) - 1)r_0$, is proportional to $r_0$ in this case, consistent with the analysis  in \cite{Noir2000}.\medskip


\textbf{Case of a large bead:} In the case of a large bead, the behavior of the system depends on whether $V_{**} \ge 0$ (Proposition 3) or $V_{**} < 0$ (Proposition 4).  In terms of the chemical potentials, these cases corresponds to whether the referential chemical potential $\mu_{R,1}$ at the outer surface of the \textcolor{blue}{solid} is not less than or less than the chemical potential $\mu_1 (= \mu_\infty)$ of the flowing free particles at that location:
$$
\mu_{R,1} \ge \mu_\infty \quad \Leftrightarrow \quad V_{**} \ge 0; \qquad \qquad \mu_{R,1} < \mu_\infty \quad \Leftrightarrow \quad V_{**} < 0,
$$
see \eqref{eq:RA20151113-1}.

 \textbf{Subcase $\eta \to \infty,  \mu_{R,1}\geq\mu_{\infty}$:} It follows from\eqref{eq:56}, \eqref{eq:59} and \eqref{eq:31} that the limiting values of $\nu(\eta), V_0(\eta)$ and $\mu(\eta)$ are all independent of $W$.  In this regime the stress build-up plays no significant role, and what limits \textcolor{blue}{accretion} is the available supply of free particles flowing from the outer surface. Following \cite{Noir2000}, we refer to this as the \emph{diffusion-limited regime}.  The estimate \eqref{eq:541121}  for the thickness in this regime can be written in terms of the chemical potentials as:
\begin{equation}\label{eq:49}
  \frac{d}{r_0} \simeq \ell_* \frac{\mu_\infty-\mu_*}{\mu_{R,1}-\mu_{\infty}} \ \frac{1}{\eta}.
\end{equation}

\textbf{Subcase $\eta \to \infty, \mu_{R,1} < \mu_{\infty}$:} \color{blue}This case, which is covered in Proposition  4, takes place when $\eta\to\infty$ with $\mu_{R,1}<\mu_{\infty}$.  

According to Proposition 4, if the radius of the bead tends to infinity, the thickness of the solid tends to infinity as well, since the ratio $\nu=r_1/r_0$ tends to $\nu_{**}>1$. To explain this result, we propose the following argument. 

When the radius of the bead tends to infinity, mechanical effects become negligibly small: indeed, when the layers that comprise the body grow on a flat surface, they can be pushed away without suffering circumferential stretch. Thus, we may think of an infinitely large bead as the same as a finite bead with vanishingly small energy $W(\lambda)$. Now, we know that treadmilling can be attained only when $V_1$ is negative. However, the kinetic equation \eqref{eq:36} tells us that if $W(\lambda)$ is small and $\mu_{R,1}<\mu_{\infty}$, then $V_1$ is positive unless the ratio $r_1/r_0$ is very large, that is to say, the body has a very large thickness. 
\color{black}

\begin{figure}[h]
\begin{center}
    {\includegraphics[width=0.6\textwidth]{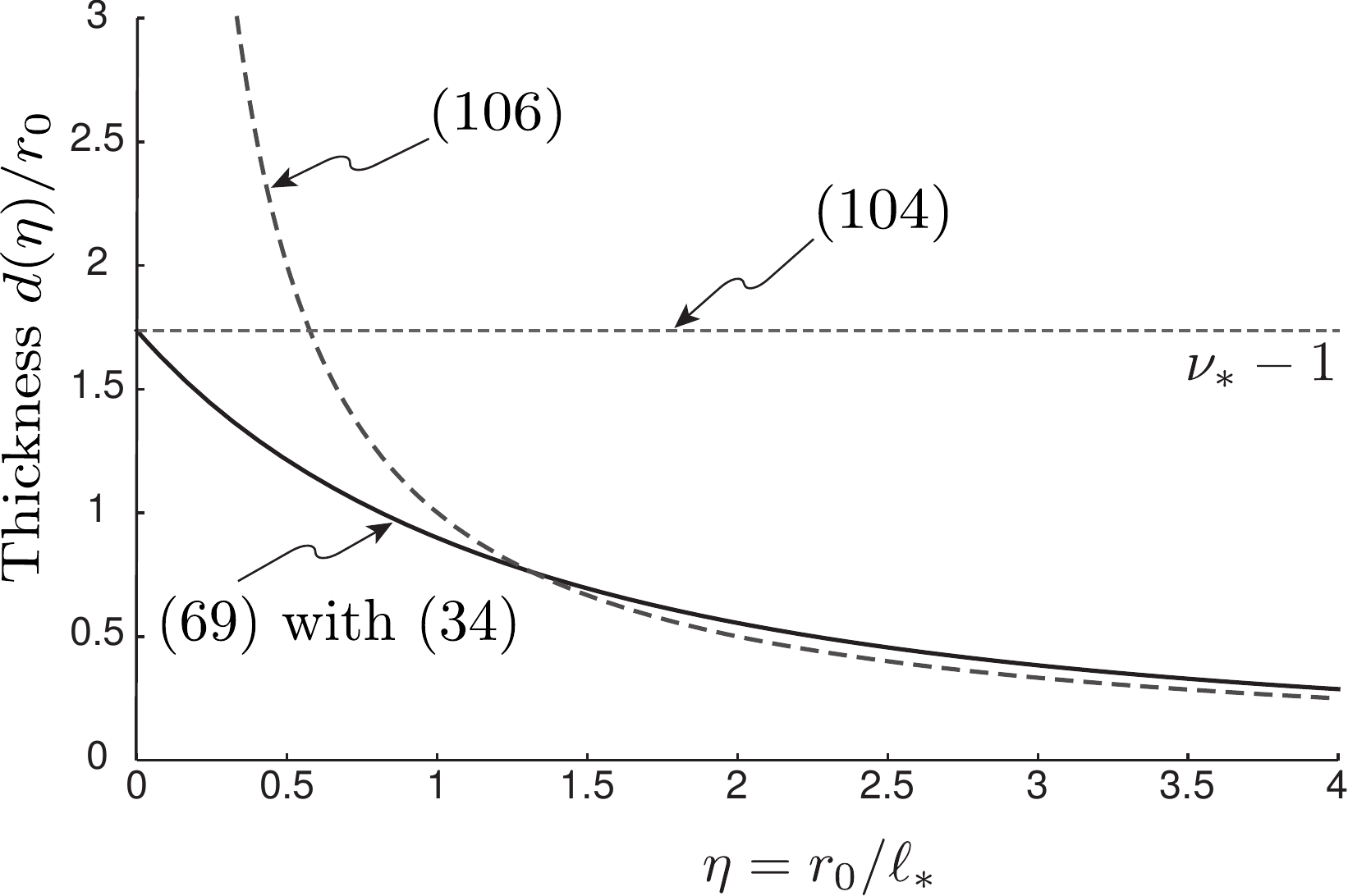}}
\end{center}
\caption{Thickness of the {solid} $d(\eta)/r_0$ versus bead radius $\eta = r_0/\ell_*$ for a neo-Hookean strain energy. Solid curve: exact solution based on numerically solving the treadmilling system \eqref{eq:55} with the neo-Hookean strain energy \eqref{eq:32}. Dashed curve: approximate formula \eqref{eq:49} in the diffusion-limited regime.  Dotted line: approximate formula \eqref{eq:63} in the stress-limited regime.} \label{fig:20151214-1}
\end{figure}

Finally we return to the case of an arbitrary value of $\eta$.  Figure \ref{fig:20151214-1} shows how the thickness of the \textcolor{blue}{solid} $d$ varies with the bead radius $r_0$ in the case of a neo-Hookean material.  The solid curve was determined by (numerically) solving the treadmilling system \eqref{eq:55} with  the neo-Hookean energy \eqref{eq:32}. The dashed curve is based on the approximate formula \eqref{eq:49} in the diffusion-limited regime.  There is good agreement for large values of  $\eta$.  The dotted straight line is based on the approximate estimation  \eqref{eq:63} in the stress-limited regime.  The two solutions agree at $\eta=0$.

\medskip 

\subsection{Summary.}

Inspired by experiments on actin motility,  we have considered a model problem which features (surface) \emph{accretion} of an annular spherical on a rigid sphere. The process considered has the unusual characteristic that new material is added to the solid, not at its outer surface, but rather at its \emph{inner surface} where it is in contact with the spherical support. 
The process of accretion is sustained by a diffusive flow of particles  both inside and outside of the solid, with particles attaching to the solid when they reach the inner surface. 
Simultaneously, particles detach from the solid at the outer surface and are returned to the particle flow. In the \emph{treadmilling} regime these two processes are balanced and the region occupied by the solid is time independent even though particles continue to attach and detach from the solid at, respectively, its the inner and outer surface.  

In order to distinguish {accretion} from motion, we found it convenient 
to choose an \emph{evolving reference configuration} that 
allowed us to label the individual material points that comprise the solid, and to keep track of the addition and removal of material points. This was achieved by the selection of a four-dimensional reference space.

The {accretion} rate is determined by three factors: the difference in the chemical potential of a particle when it is free and when it is attached to the solid; the strain energy of the {solid}; and the radial normal stress. The {\it driving force for accretion} that we
derive in a thermodynamically consistent manner involves all three of these factors. On the assumption of \emph{small deviations from thermodynamical equilibrium} we take a linear kinetic relation between the driving force and the accretion rate.

We have established necessary and sufficient conditions under which a treadmilling state exists, 
and we have shown that, when those conditions hold, the solution is unique.  Moreover according to our model the build-up of stress at the inner boundary hinders accretion,  whereas at the outer surface, it is the build-up of strain energy, not stress,  that causes the ablation rate at the outer surface to increase. These results apply to \emph{arbitrary} uniform, isotropic, incompressible, elastic materials.

By an asymptotic analysis we have shown that for small beads the thickness of the solid is proportional to the radius of the support and is strongly affected by the stiffness of the solid, whereas for large beads the stiffness of the solid is essentially irrelevant, and the thickness is proportional to a characteristic length that depends on the parameters that govern diffusion and {accretion} kinetics.

\section{Acknowledgments}
TC wishes to gratefully acknowledge the support of the MIT-Technion Post-Doctoral Fellowship
Program. GT thanks  Antonio De Simone and Paolo Podio-Guidugli for fruitful discussions, and the Italian INdAM-GNFM for financial support through the initiative ``Progetto Giovani''.

\color{red}{ 


 





\color{black}

\bibliographystyle{abbrv}


\end{document}